\documentclass[referee]{aa}

\usepackage[dvips]{graphics}
\usepackage[dvips]{graphicx}

\begin{document}

\thesaurus{07(06.01.2; 06.06.1; 06.19.3; 08.01.1; 08.01.2; 08.12.1)}

\title{The effect of stellar activity on the Li\,{\sc i} 6708, Na\,{\sc i} 5896
and K\,{\sc i} 7699 \AA\ lines.}

\subtitle{A comparison with the Pleiades, field stars and the 
Sun\thanks{Based on
 observations collected with the Gregory-Coud\'e telescope, operated on the
 island of Tenerife by the Universit\"ats-Sternwarte G\"ottingen in the Spanish
 Observatorio del Teide of the Instituto de Astrof\'{\i}sica de Canarias.}}

\author{D.~Barrado y Navascu\'es \inst{1} 
         \and R.J.~Garc\'{\i}a L\'opez \inst{2,3}
         \and G. Severino \inst{4}
	 \and M.T.~Gomez \inst{4}
	 }

\offprints{D.~Barrado y Navascu\'es}

 \institute{Departamento de F\'\i sica Te\'orica, C-XI-506, 
 Universidad Aut\'onoma de Madrid, Cantoblanco, E-28049 Madrid, 
 Spain\\e-mail: barrado@pollux.ft.uam.es
 \and
 Instituto de Astrof\'{\i}sica de Canarias, E-38200 La Laguna, Tenerife, 
 Spain\\e-mail: rgl@ll.iac.es
 \and 
 Departamento de Astrof\'\i sica, Universidad de La Laguna, Av. 
 Astrof\'\i sico Francisco S\'anchez s/n, E-38071 La Laguna, Tenerife, Spain
 \and
 Osservatorio Astronomico di Capodimonte, Via Moiariello, 16, I-80131 Napoli,
 Italy\\e-mail: 
severino@na.astro.it, gomez@na.astro.it
 }                     
  
\date{Received date, 9 June 2000; accepted date, 12 March 2001 }
\titlerunning{The effect of stellar activity on alkali lines}
\authorrunning{D.~Barrado y Navascu\'es et al.}
\maketitle

\begin{abstract} 
An analytical model has been developed to empirically study the effects of
stellar spots and faculae on the observed equivalent
widths of   
Li\,{\sc i} 6708, Na\,{\sc i} 5896  and K\,{\sc i} 7699 \AA\ lines 
(and abundances in the case of lithium) 
in late-type stars, taking into account
the changes in the observed magnitudes and colors.
Solar spectra corresponding to
different active regions are used as input data and a range of filling factors
are applied to simulate the surfaces of stars with different levels of
activity. Detailed comparisons between predicted and observed photometric
colors and equivalent widths are made for late-type stars of the Pleiades and the field. The
observed dispersions in K\,{\sc i} and Li\,{\sc i} equivalent widths for Pleiades stars can be
partially accounted by the simultaneous effects of activity on colors and the
line formation, indicating that the lithium-rotation connection suggested for
$\sim 0.7-0.9$ $M_\odot$ Pleiades stars could be due in part to the stellar
activity. However, under realistic values for the filling factors, only a
small portion of the observed spread could be explained by these effects.


\end{abstract}

\section{Introduction}
\label{sec1}

The formation of  neutral alkali resonance lines in the solar photosphere is
known to be affected by the presence of active regions and spots on the solar
surface, because the fractions of neutral alkali atoms in their ground states
are  very sensitive to the photospheric temperature structure (see e.g. 
Giampapa 1984; Caccin et al. 1993; Severino et al. 1994, and references
therein). This is expected also to be the case for the photospheres of other
late-type (F- to early M-type) stars with similar atmospheric characteristics.
Different authors have dealt with the comparison of observed and synthetic
spectra of Na\,{\sc i} D ($5890-5896$ \AA) and K\,{\sc i} 7699 \AA\ lines in
stars of different spectral types (Covino et al. 1993; Andretta et al. 1997;
Tripicchio et al. 1997, 1999; Short \& Doyle 1998), showing the benefits and
difficulties in using the lines of neutral alkali to probe the effect of
activity on the photospheres of these stars. 

Both the Li\,{\sc i} 6708 and K\,{\sc i} 7699 \AA\ features are resonance lines
with similar excitation potentials and similar formation conditions: any line
formation effect which affects the K\,{\sc i} line should also affect the
Li\,{\sc i} line. In contrast with potassium, lithium is destroyed inside the
stellar interior by $(p,\alpha)$ nuclear reactions in those layers where the
temperature reaches $\sim 2.5\times 10^6$ K. The K\,{\sc i} line appears then
as a key tool for identifying and isolating the possible effects of activity
on the observed Li abundances in late-type stars, since its surface abundance
does not depend on stellar age, rotational velocity, etc.
 
Open clusters, due to the common properties shared by the stars belonging to
them (same age and metallicity), provide excellent opportunities to study the
details of stellar structure and evolution. Lithium abundances have been
derived for a significant number of clusters and different mixing mechanisms
have been invoked to explain the observed patterns (see recent reviews by
Balachandran 1995; Pinsonneault 1997; Mart\'{\i}n 2001; Vauclair 2001). 
These observations indicate that the abundance  decreases with
temperature and   stellar age, and that there is a 
star-to-star scatter of the Li abundance
 for the same temperature.
In particular, the existence of a lithium-rotation
 connection has been suggested
for $\sim 0.7-0.9$ $M_\odot$ stars in young clusters such as the Pleiades
(Butler et al. 1987; Garc\'\i a L\'opez et al. 1991a,b; Soderblom et al. 1993;
Garc\'\i a L\'opez et al. 1994; Jones et al. 1996) and Alpha  Persei cluster
(Balachandran et al. 1988; Randich et al. 1998), indicating that fast rotators,
as a group, show  higher lithium abundances
 than slow rotators. Balachandran et al. (1998)
discuss different aspects of the lithium-rotation connection and mixing in
late-type stars. Note, however, that slightly older clusters do not 
clearly show this connection, as  in the case of M35 (Barrado y
Navascu\'es et al. 2001b). Alternatively, it has also been suggested that
chromospheric activity could affect the formation of the Li line, partially
accounting for the observed dispersion (Houdebine \& Doyle 1995; Barrado y
Navascu\'es 1996; Russel 1996; Jeffries 1999; King et al. 2000). In this case
the Li-rotation connection would be a Li-activity-rotation relation (higher
rotation means higher activity) and the observed dispersion in Li equivalent
widths (Ws) would not (or only in part) correspond to a real dispersion in Li
abundances. 

Soderblom et al. (1993) found that the K\,{\sc i} 7699 \AA{ }
equivalent widths of K-type
stars in the Pleiades exhibited a similar trend to that of Li\,{\sc i} 6708
 \AA{ } but with a
lower spread. They concluded in their study that activity could only partially
influence the observed Li spread. Stuik et al. (1997) compared observed and
synthetic Li\,{\sc i} and K\,{\sc i} spectra (affected by the presence of spot 
and plages) for late-type Pleiades stars. They found that these lines are not
sensitive to the presence of a chromosphere, and that they respond to the
effects of activity on the stratification in the deep photosphere. However,
they could not establish whether or not magnetic activity is the major
contributor to the observed K\,{\sc i} (and Li\,{\sc i}) scatter. More recently,
King et al. (2000) addressed this problem by using differential Li abundances
(with respect to fitted values vs. effective temperature) and differential
K\,{\sc i} equivalent widths. They invoked an incomplete treatment of line
formation (in particular, the role of ionization in reducing the the Li\,{\sc
i} line optical depth; Houdebine \& Doyle 1995) to explain the correlations
found between differential Li and K values and differences in stellar
activity (measured using the Ca\,{\sc ii} IR triplet).    

None of these previous works has clearly established the quantitative effects
of stellar activity on the K\,{\sc i} and Li\,{\sc i} lines. In this paper, we
have adopted an empirical approach to this study and included the Na\,{\sc i}
5896 \AA\ line as a complementary indicator. An analytical model has been
developed to evaluate the simultaneous effect of surface inhomogeneities on the
observed line equivalent widths and stellar photometric colors. The inputs to
this model come from high-quality, high-resolution observations carried out on
different regions of the solar disk. Section 2 briefly describes the
observational material, while Section 3 provides the details of the model. A
comparison between model predictions and photometric data for the Pleiades is
shown in Section 4. Section 5 discusses the effect of stellar spots  and
faculae on the alkali and compares these results with K\,{\sc i} and Li\,{\sc
i} measurements in late-type Pleiades stars. Finally, Section 6 summarizes our
main conclusions.

\begin{figure}
\resizebox{8.7cm}{!}{\includegraphics{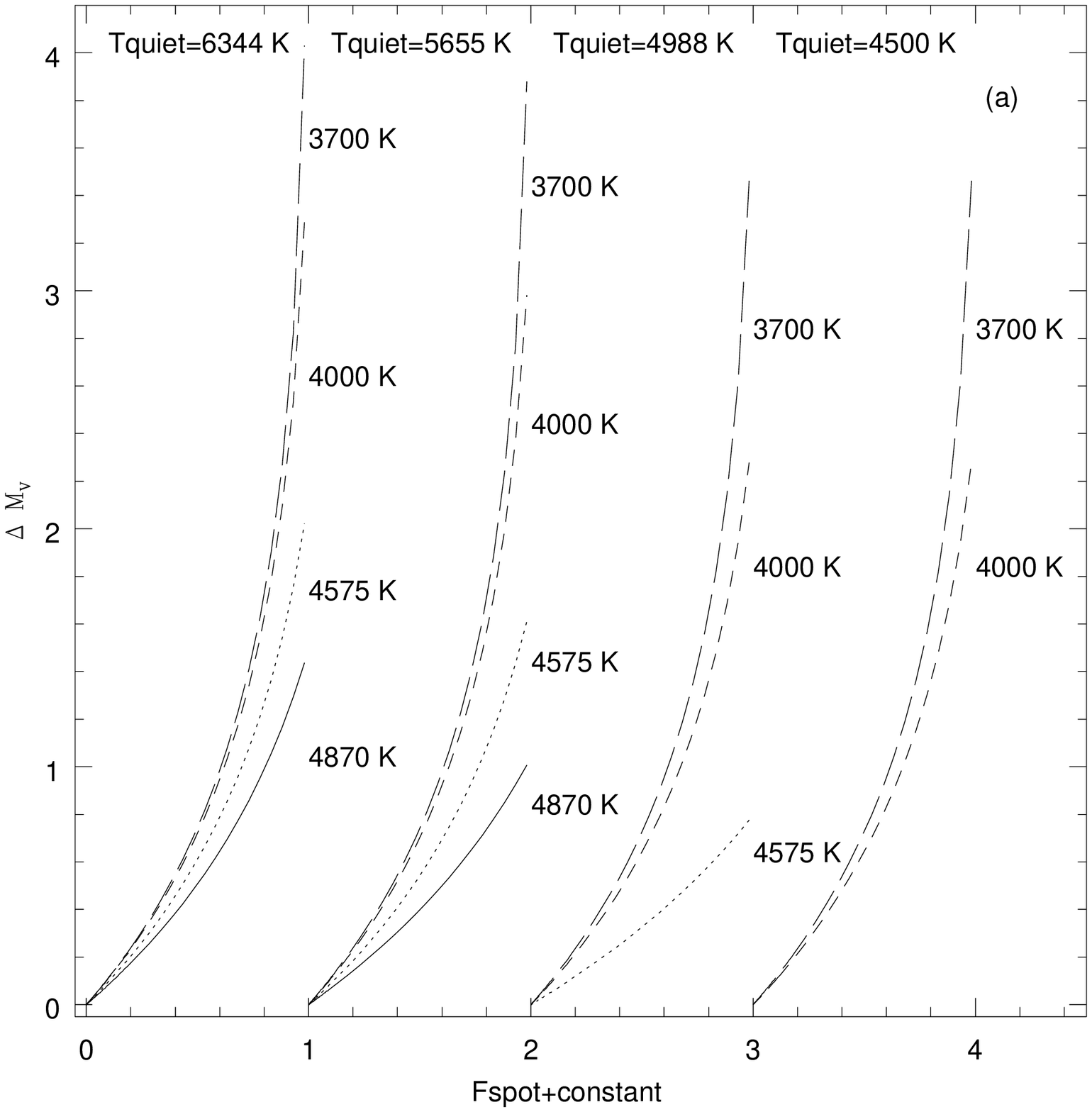}}
\resizebox{8.7cm}{!}{\includegraphics{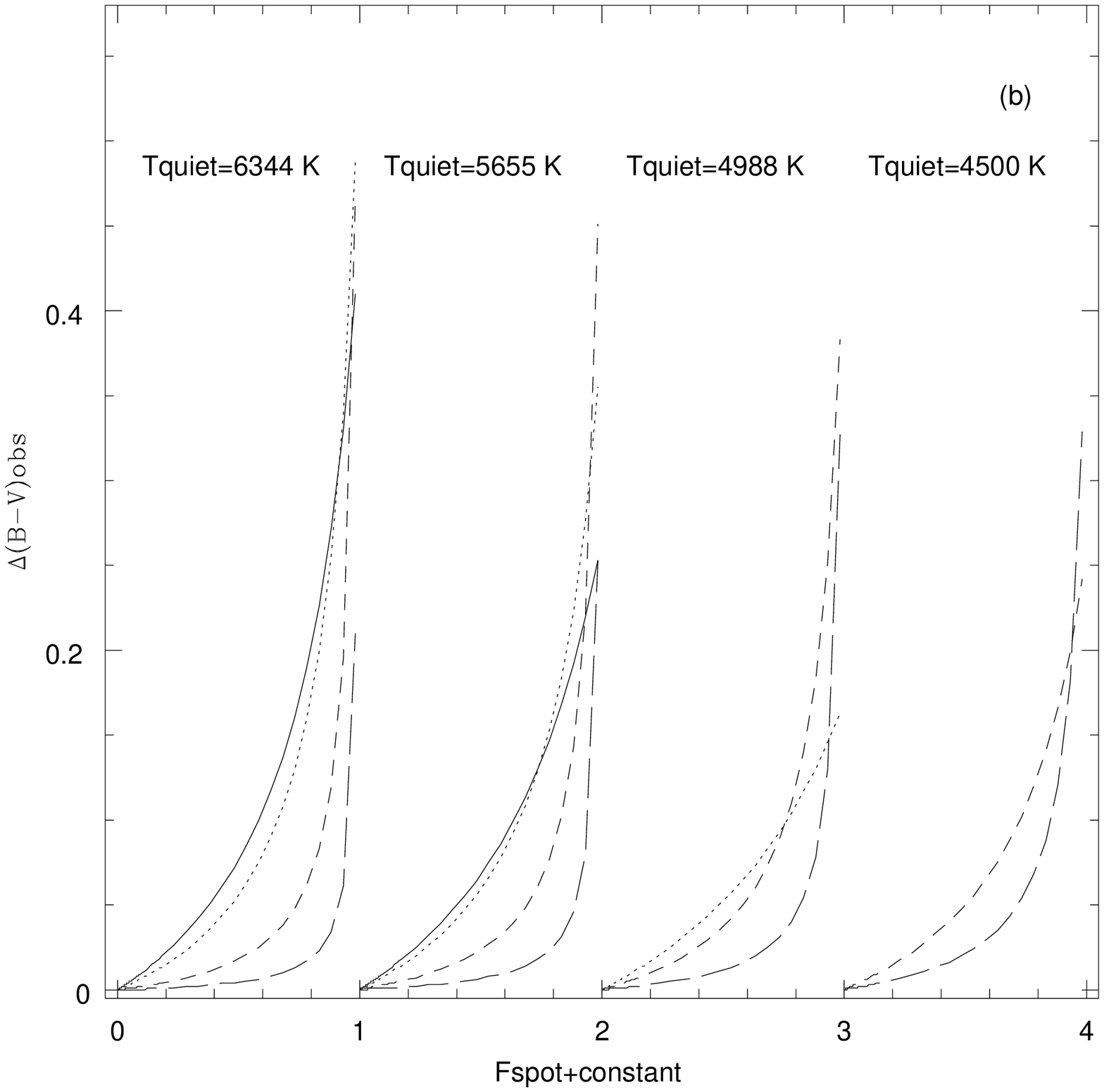}}
\caption[]{\label{fig1} Results of our model for a star with
 T$_{\rm quiet}$=6344, 5655, 4988 and 4500 K. The solid line shows the computation for a spot at 4870 K,
the dotted line  represents the values for a spot at 4575 K,
the short dashed line corresponds  to 4000K.  Finally, the long dashed line 
represents the computation for a spot at 3700 K.
{\sl (a).} \ Variation of the visual magnitude, $\Delta$M$_V$, against the
filling factor of the spots, F$_{\rm spot}$. For clarity, we have added one, two  and three units
to the models having T$_{\rm quiet}$ equal to 5655, 4988 and 4500 K, respectively.
  {\sl (b).} \
Dependence of variation of the observed color index  (B--V) on the spot filling factor.  }
\end{figure}

\setcounter{table}{0}

\begin{table*}
\caption[ ]{Alkali equivalent widths measured on different solar regions.}
\begin{tabular}{lccccc}
\hline 
  Line  	   & \multicolumn{5}{c}{Equivalent width  (m\AA)  }        \\
\cline{2-6}							       
		   & Quiet             & Faculae           &   Average Umbra   &    Central Umbra  &  Spot 2      \\
  (\AA)	  	   &                   &                   &                   &                   &              \\
\hline
Li\,{\sc I}  6708  &\phantom{16} 3.977 & \phantom{16}3.569 & \phantom{2}20.247 & \phantom{3}49.080 &   --         \\
K\,\,{\sc I} 7699  & 179.5             &            177.3  &     294.5         &           373.4   & 1139.0$^{(1)}$\\ 
Na\,{\sc I}  5896  & 599.0             &            531.4  &    1118.4         &          1786.6   & 5180.9$^{(1)}$\\
Temperature  (K)   & 5655              &            --     &    4870           &          4575     & 3700$^{(1)}$\\   
\hline
\end{tabular}
$\,$\\
$^{(1)}$ From Brynildsen et al. (private comm.), after Tripicchio et al. (1997, 1999).
\end{table*}

\section{Observational material}
\label{sec2}

The observations used in this work were carried out in September 1994, using
the echelle spectrograph of the Gregory-Coud\'e telescope located at the Teide
Observatory, Tenerife (Spain). This instrument provides a great stability
together with very high spectral resolution. Spectra with a resolving power of
$\lambda/\Delta\lambda\sim 3.5\times 10^5$ were recorded in each spectral range
using a 1024$\times$1024 pixel Thomson CCD (19$\mu$m~pixel$^{-1}$). Long slit
observations were performed in order to obtain a spatial coverage of 
160\arcsec{ } in a given measurement. Each final spectrum is typically the
average of 20 independent images (50 in the case of Li), which were reduced
using MIDAS following the usual procedure for solar observations carried out
with this instrumentation. Final signal-to-noise ratios are in the range
$1000-4000$. Additional details on the observations and data reduction can be
found in Barrado y Navascu\'es et al. (1995) and Barrado y Navascu\'es (1996).
Table 1 shows the equivalent widths measured for different locations of the
observed active region and its surroundings. Since our original spectra have
very high resolution, we were able to  remove either
the nearby TiO and CN bands --which appear at low temperatures-- in the case of
Li\,{\sc i}, or the weak lines in the case of K\,{\sc i} and Na\,{\sc i}.  
These equivalent widths  will be used as the input data for the model proposed in the next
section.
Additional measurements (for Na and K) were selected from Tripicchio 
et al. (1997, 1999)  for a sunspot at 3700 K.

\begin{figure}
\resizebox{8.7cm}{!}{\includegraphics{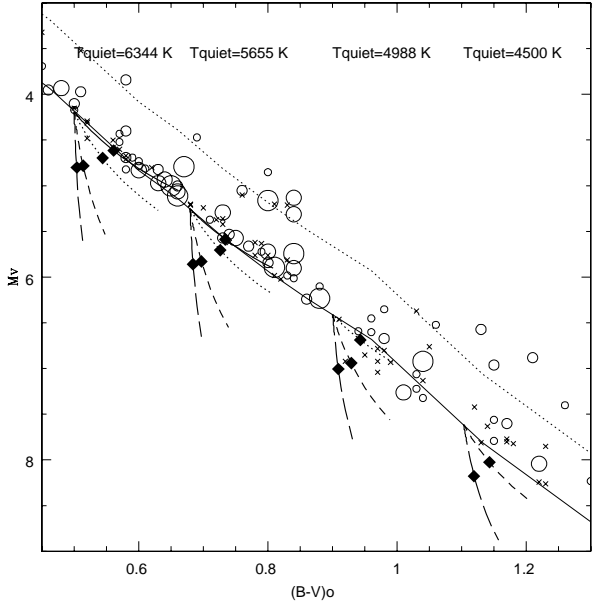}}
\caption[]{\label{fig2} Visual absolute magnitude against the color index
(B--V) for  four cases presented in this study (thin-solid, dotted, short dashed
and long dashed lines for spots with 4870, 4575,  4000 and 3700 K, respectively).
 We have computed the observed
photometry for stars with T$_{\rm quiet}$=6344, 5655, 4988 and 4500 K, corresponding
to (B--V)$=0.50$, 0.68, 0.90 and 1.10, respectively. The MS appears as a thick solid
line, whereas the MS for twin stars is represented as a short dashed line
(MS--0.75$^{\rm mag}$). Pleiades data are displayed as open circles and the
size increases with increasing stellar activity. Star without known activity
appear as plus symbols. Solid diamonds represent our
calculations for F$_{\rm spot}$=0.50. All calculations  are displayed up to 
F$_{\rm spot}$=0.80.}
\end{figure}

\section{The model}
\label{sec3}

We estimate in this section the behavior of the equivalent widths of Na\,{\sc
i} 5896, Li\,{\sc i} 6708 and K\,{\sc i} 7699 \AA\ lines associated with the
presence of active regions in late-type stars. Stellar spots (and other surface
inhomogeneities) affect the observed photometry due to the lower luminosity of
the spot in comparison with that of a quiet region of equivalent size. On the
other hand, it is known that the equivalent widths (Ws) of the alkali show
different values in different solar regions (as  can be seen in Table 1) 
associated with the different temperatures of the line forming regions. A
simple mathematical model is used here to compute an ``observed''  (predicted)
color index --in our case, (B-V)-- and the corresponding effective temperature,
as well as the ``observed'' (predicted) equivalent widths  of the alkali
lines for a given set of parameters. This model is intended to reproduce what
happens in late-type stars from our knowledge of the solar surface.

The computation starts from a fictitious star free of active regions and 
having a given effective temperature. Then, different active regions (spots and
faculae) are added to its photosphere in a similar fashion  to that followed by
Pallavicini et al. (1993).

\subsection{Computing the photometry}
\label{sec3.1}

Estimating the observed photometry for a stellar disk which contains spots
requires the previous computation of the contribution of the region covered by
them, i.e. its filling factor, F$_{\rm spot}$. In a similar way, we should
consider the fraction covered by faculae, F$_{\rm faculae}$, and the fraction
filled with quiet regions. For simplicity, we have supposed that the
photometric properties of these two last regions are identical. Therefore, it
is only necessary to take into account F$_{\rm spot}$ and \{1--F$_{\rm
spot}$\}. We have not considered the effect of limb darkening either.

The absolute magnitude of the region covered by spots can be expressed by:

\begin{equation}
{\rm M_{V,spot} = M_{V,star}^{(B-V)_{spot}} 
- 2.5 \log \{F_{spot}\times 
[R_{(B-V)_{quiet}}/R_{(B-V)_{spot}}]^2 \}     },
\end{equation}

\noindent where M$_{\rm V,star}^{(B-V)_{\rm spot}}$ is the magnitude of a star
with color index (B--V) equal to that one of the spot,  R$_{\rm (B-V)_{\rm
quiet}}$ is the radius of a star of color equal to the quite region (the real
radius of the considered star), and R$_{\rm (B-V)_{\rm spot}}$ is the radius of
a star which would have the same color of the spot.

In the case of the region covered by faculae and the quiet photosphere:
 
\begin{equation}
{\rm  M_{V,faculae} = M_{V,star}^{(B-V)_{faculae}} - 2.5
\log F_{faculae} },
\end{equation}

\noindent and

\begin{equation}
{\rm  M_{V,quiet} = M_{V,star}^{(B-V)_{quiet}} 
- 2.5 \log \{1-F_{spot}-F_{faculae} \}  },
\end{equation}

\noindent where we have assumed that ${\rm M_{V,star}^{(B-V)_{faculae}}}$ is
equal to ${\rm M_{V,star}^{(B-V)_{quiet}}}$. In a similar way it is possible to
obtain the absolute magnitude in the B band --M$_{\rm B}$-- for the regions
covered by spots, faculae and quiet region, and derive the color indices for 
each region.

Finally, the observed magnitudes can be predicted using the expression:

\begin{equation}
{\rm M_V^{obs} = -2.5 \log \{10^{-0.4\times M_{V,spot}} +
10^{-0.4\times M_{V,faculae}} + 10^{-0.4\times M_{V,quiet}} \}.  
}
\end{equation}
 
An equivalent expression gives M$_{\rm B}^{\rm obs}$, allowing the computation
of the color index (B--V)$_{\rm obs}$. The Appendix A contains a detailed
justification of these expressions. Some results are shown in Figure 1.

\subsection{Computing the equivalent widths}
\label{sec3.2}

There are several factors which affect the observed equivalent widths. They
depend, among other parameters, on the filling factors of spots and faculae.
Moreover, the Ws depend on the local continuum level. Therefore, it is
necessary to compute the ratios  $\alpha$$_{\rm line}$, between the fluxes from
the region covered by spots and the quiet region (and faculae) at the studied
wavelengths --$\alpha$$_{\rm line}$=Flux(spot)/Flux(quiet):

\begin{equation}
{\rm  \alpha_{line} =
 { e^{(K(\lambda)/T_{quiet})}-1
 \over
e^{(K(\lambda)/T_{spot})}-1} },
\end{equation}

\noindent where  and K($\lambda$) is equal to 
2.1443$\times$10$^4$ for  Li\,{\sc i} 6708, 
2.4408$\times$10$^4$ for  Na\,{\sc i} 5896, and 
1.8683$\times$10$^4$ for  K\,{\sc i} 6799.
Then, the equivalent width for a particular
 line is estimated using the
following expression (see Appendix B):

\begin{equation}
{\rm  W_{obs}^{line} = {
\alpha_{line}~F_{spot}~W^{line}_{spot} +
F_{faculae}~W^{line}_{faculae} +
F_{quiet}~W^{line}_{quiet} \over
\alpha_{line}\times~F_{spot}
+ F_{faculae} + [1 - F_{spot} - F_{faculae}]     } },
\end{equation}

\noindent where W$_{\rm obs}^{\rm line}$ is the predicted equivalent width for
a certain line and  W$_{\rm spot}^{\rm line}$, W$_{\rm faculae}^{\rm line}$
and W$_{\rm quiet}^{\rm line}$ are the actual equivalent widths of the spot,
faculae and quiet region, respectively.

\begin{figure}
\resizebox{7.cm}{!}{\includegraphics{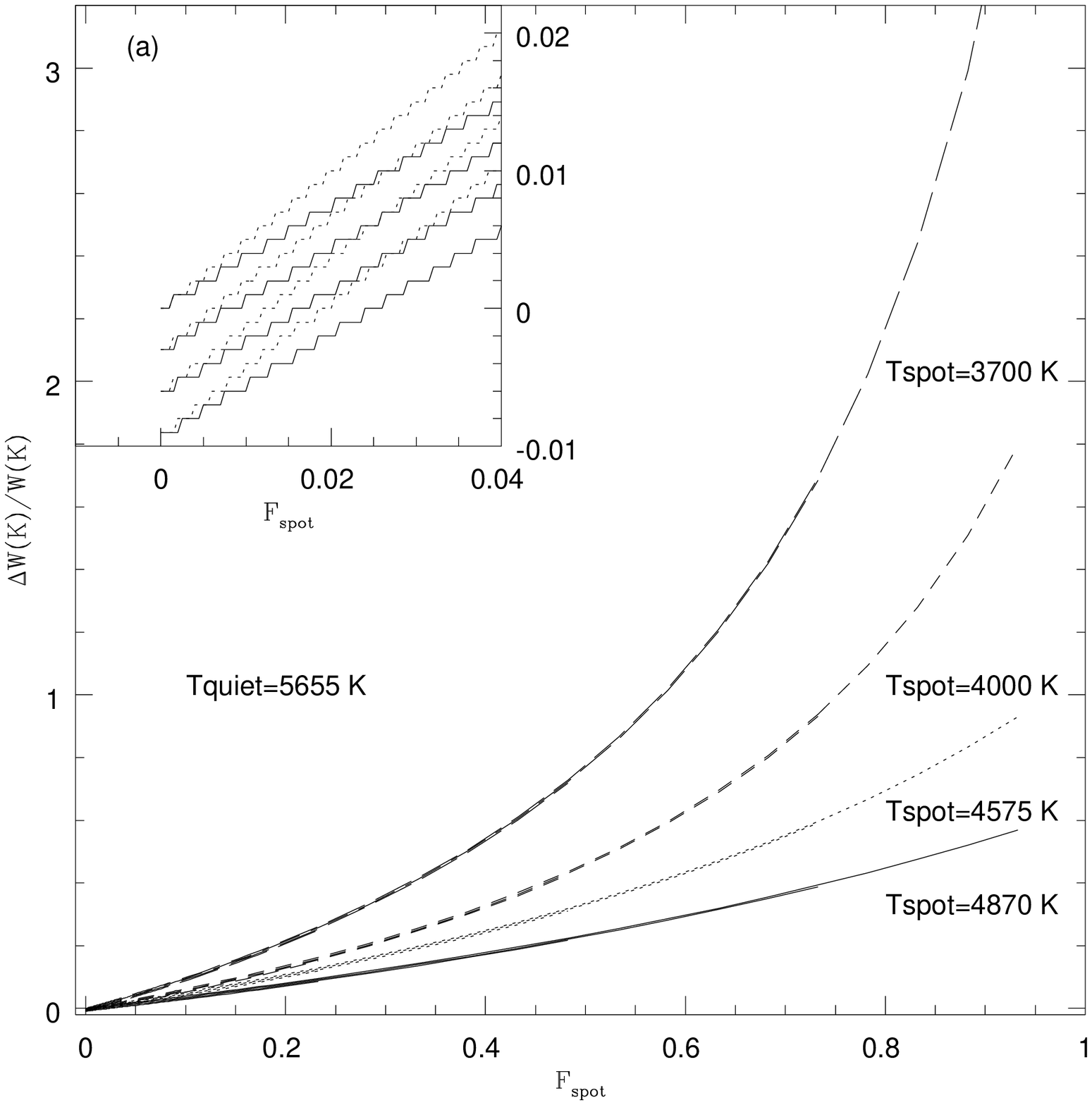}}
\resizebox{7.cm}{!}{\includegraphics{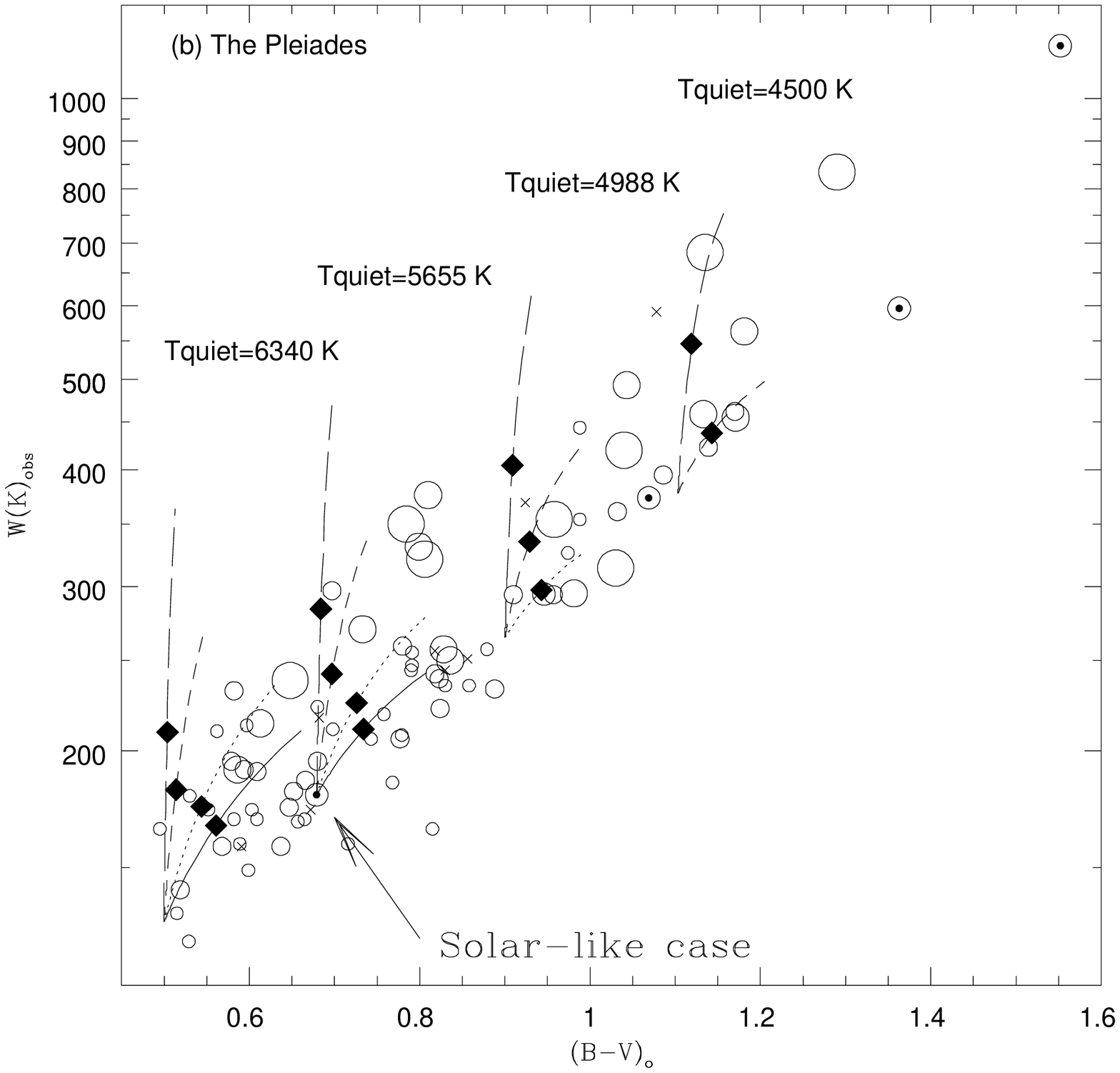}}
\resizebox{7.cm}{!}{\includegraphics{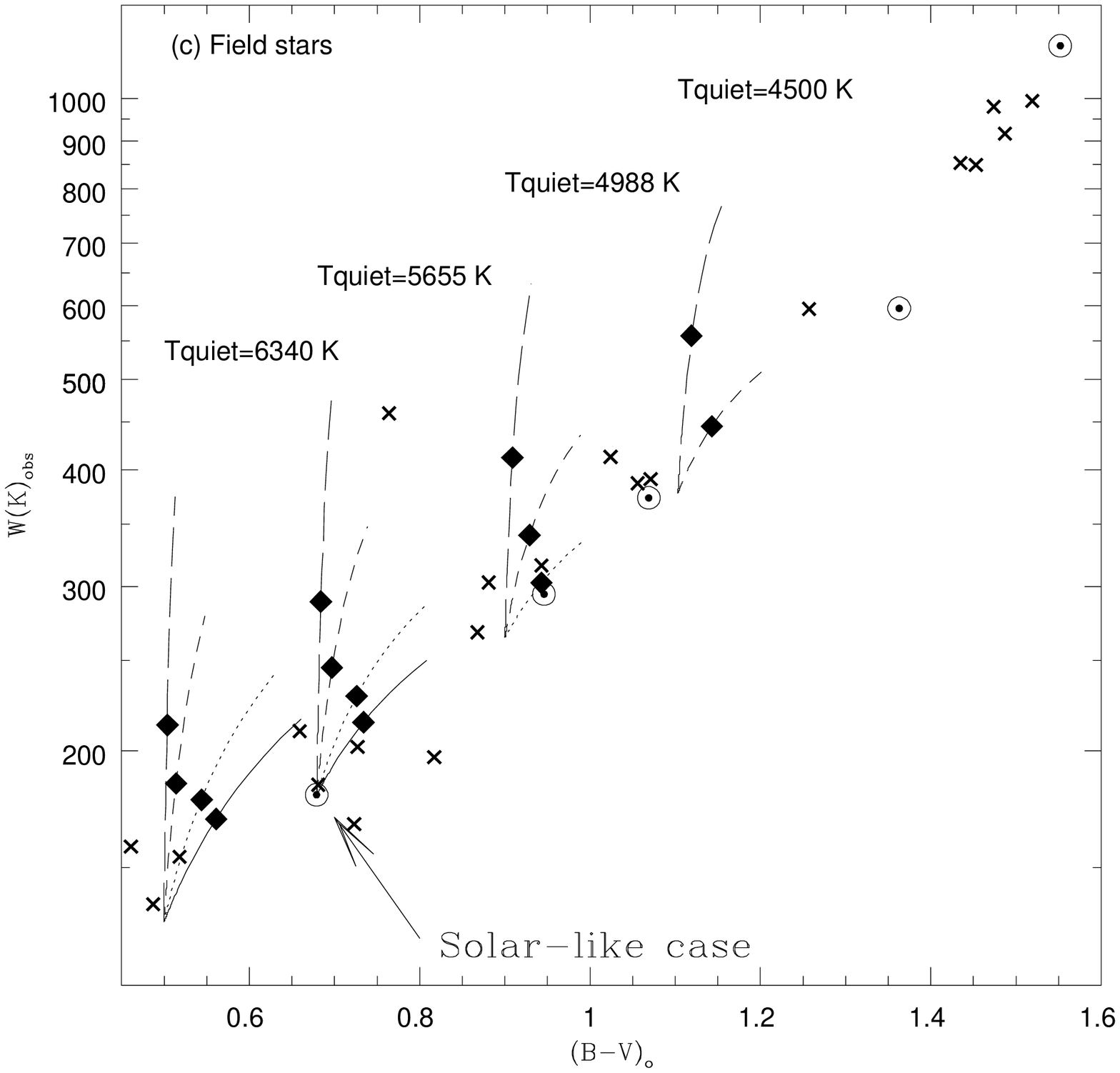}}
\caption[]{\label{fig3} The effect on potassium.
{\sl (a)} 
\ Relative variation of the observed potassium
equivalent width  against the filling factor of the spot. The set displayed
with solid lines were computed with T$_{\rm spot}$=4870 K, whereas the set
plotted with dotted, short dashed and long dashed lines corresponds to T$_{\rm spot}$=4575 K,
T$_{\rm spot}$=4000 K, and  T$_{\rm spot}$=3700 K
For a given
set, we computed our results using different faculae filling factors. From top
to bottom, F$_{\rm faculae}$=0.00, 0.25, 0.50 and 0.75.
{\sl (b)} \ Comparison with Pleiades data.
 Symbols as in  Figure 2.
Values measured/estimated for different sunspots appear with the solar symbol (dot$+$circle).
{\sl (c)} Comparison with field stars (from Tripicchio et al. 1999). These main sequence stars
are represented as crosses. }
\end{figure}

\subsection{Effective temperatures, radii, colors, and magnitudes}
\label{sec3.3}

Tripicchio et al. (1999) have shown that the potassium equivalent
width at  7699 \AA{ } is an excellent temperature indicator
(in a very well defined spectral range).
 Therefore, we have computed 
the effective temperatures from their expression:

\begin{equation}
{\rm  Log (W(K)/7699) = \\
{ -2.85 + 3.29 \times 10^{-5} \times Teff - 
1.43 \times 10^{-7} \times Teff^2 + 1.11 \times 10^{-11} \times Teff } },
\end{equation}

Color indices (B--V) were derived from  the Thorburn et al. (1993)
 scale temperature,   namely:

\begin{equation}
{\rm  Teff =
{ 5040
 \over
 0.5247+0.5396 \times (B-V) } },
\end{equation}

Stellar radii were estimated from (B-V) colors, after
Schmidt-Kaler (1982).

Absolutes magnitudes were derived using a Pleiades empirical
main sequence (MS, see Figure 2), with a color excess of E(B-V)=0.04
and a distance modulus of (m-M)=5.728. Other examples of empirical
main sequences for the Pleiades, Alpha Persei cluster, M35 and field
stars can be found in Bouvier et al. (1998), Stauffer et al. (1999),
Barrado y Navascu\'es et al. (2001b) and
Barrado y Navascu\'es et al. (1999), respectively.

\setcounter{table}{1}
\begin{table*}
\caption[ ]{Input model parameters for Pleiades and field stars.
Columns  \#4 and \#6 provide the (B-V) and radius
 of a star with the temperature of the spot.}
\begin{tabular}{cccccccccc}
\hline 
Model  & M$_V$  &
\multicolumn{2}{c}{(B-V)$_0$}                   &\,&
\multicolumn{2}{c}{Radii (R$_\odot$)}           &\,&
\multicolumn{2}{c}{Temperature (K)}             \\
\cline{3-4}  \cline{6-7} \cline{9-10}  
     & Quiet & Quiet & Spot  && Quiet & Spot  &&  Quiet & Spot  \\
(1)  & (2)   & (3)   & (4)   && (5)   &  (6)  &&  (7)   &  (8)  \\
\hline                        		         	       
 1a  & 4.192 & 0.500 & 0.946 && 1.218 &0.794  &&  6344  & 4870  \\
 1b  & 4.192 & 0.500 & 1.069 && 1.218 &0.750  &&  6344  & 4575  \\
 1c  & 4.192 & 0.500 & 1.363 && 1.218 &0.630  &&  6344  & 4000  \\  
 1d  & 4.192 & 0.500 & 1.552 && 1.218 &0.408  &&  6344  & 3700  \\  
		                   	   		       
 2a  & 5.253 & 0.679 & 0.946 && 0.921 &0.794  &&  5655  & 4870  \\
 2b  & 5.253 & 0.679 & 1.069 && 0.921 &0.750  &&  5655  & 4575  \\
 2c  & 5.253 & 0.679 & 1.363 && 0.921 &0.630  &&  5655  & 4000  \\
 2d  & 5.253 & 0.679 & 1.552 && 0.921 &0.408  &&  5655  & 3700  \\
		                   	   		       
 3a  & 6.411 & 0.900 & 0.946 && 0.811 &0.794  &&  4988  & 4870  \\
 3b  & 6.411 & 0.900 & 1.069 && 0.811 &0.750  &&  4988  & 4575  \\
 3c  & 6.411 & 0.900 & 1.363 && 0.811 &0.630  &&  4988  & 4000  \\
 3d  & 6.411 & 0.900 & 1.552 && 0.811 &0.408  &&  4988  & 3700  \\
		                   	   		       
 4c  & 7.612 & 1.103 & 1.363 && 0.737 &0.630  &&  4500  & 4000  \\
 4d  & 7.612 & 1.103 & 1.552 && 0.737 &0.408  &&  4500  & 3700  \\
\hline
\end{tabular}
$\,$\\
\end{table*}

\setcounter{table}{2}

\begin{table*}
\caption[ ]{Sodium and potassium equivalent widths for   Pleiades and field stars, in m\AA.}
\begin{tabular}{ccccccccc}
\hline 
Model  & 
\multicolumn{2}{c}{Temperature (K)}             &\,&
\multicolumn{2}{c}{W(Na\,{\sc I}5896) (m\AA)}  &\,&
\multicolumn{2}{c}{W(K\,{\sc I}7699) (m\AA)} \\
\cline{2-3}  \cline{5-6} \cline{8-9} 
     &   Quiet & Spot &&    Quiet    &    Spot     &&    Quiet    &   Spot     \\
(1)  &   (2)   &  (3) &&    (4)      &    (5)      &&    (6)      &    (7)     \\
\hline  		       
 1a  &   6344  & 4870 && 348.9$^{(1)}$ &1118.4$^{(2)}$ && 131.1$^{(5)}$ & 294.5$^{(2)}$ \\
 1b  &   6344  & 4575 && 348.9$^{(1)}$ &1786.6$^{(2)}$ && 131.1$^{(5)}$ & 373.4$^{(2)}$ \\
 1c  &   6344  & 4000 && 348.9$^{(1)}$ &3522.9$^{(3)}$ && 131.1$^{(5)}$ & 595.9$^{(5)}$ \\  
 1d  &   6344  & 3700 && 348.9$^{(1)}$ &5198.9$^{(4)}$ && 131.1$^{(5)}$ &1139.0$^{(4)}$ \\  
       	   	                                	      	           
 2a  &   5655  & 4870 &&599.0$^{(1,2)}$&1118.4$^{(2)}$ &&179.5$^{(2,5)}$& 294.5$^{(2)}$ \\
 2b  &   5655  & 4575 &&599.0$^{(1,2)}$&1786.6$^{(2)}$ &&179.5$^{(2,5)}$& 373.4$^{(2)}$ \\
 2c  &   5655  & 4000 &&599.0$^{(1,2)}$&3522.9$^{(3)}$ &&179.5$^{(2,5)}$& 595.9$^{(5)}$ \\
 2d  &   5655  & 3700 &&599.0$^{(1,2)}$&5198.9$^{(4)}$ &&179.5$^{(2,5)}$&1139.0$^{(4)}$ \\
		                   	        	      	           	              	        
 3a  &   4988  & 4870 &&1167.4$^{(1)}$ &1118.4$^{(2)}$ && 264.5$^{(5)}$ & 294.5$^{(2)}$ \\
 3b  &   4988  & 4575 &&1167.4$^{(1)}$ &1786.6$^{(2)}$ && 264.5$^{(5)}$ & 373.4$^{(2)}$ \\
 3c  &   4988  & 4000 &&1167.4$^{(1)}$ &3522.9$^{(3)}$ && 264.5$^{(5)}$ & 595.9$^{(5)}$ \\
 3d  &   4988  & 3700 &&1167.4$^{(1)}$ &5198.9$^{(4)}$ && 264.5$^{(5)}$ &1139.0$^{(4)}$ \\
      	   	                                	      	           
 4c  &   4500  & 4000 &&2154.7$^{(1)}$ &3522.9$^{(3)}$ && 377.7$^{(5)}$ & 595.9$^{(5)}$ \\
 3d  &   4500  & 3700 &&2154.7$^{(1)}$ &5198.9$^{(4)}$ && 377.7$^{(5)}$ &1139.0$^{(4)}$ \\
\hline
\end{tabular}
$\,$\\
$^{(1)}$ From W(Na\,{\sc I}5896)=0.473$\times$10$^{(-0.788+1.311(B-V))}$, adapted from Tripicchio et al. (1997).\\
$^{(2)}$ Measured. See Table 1.\\
$^{(3)}$ From field stars. See Figure 3c.\\
$^{(4)}$ From direct measurement. See Tripicchio et al. (1999).\\
$^{(5)}$  From W(K\,{\sc I}7699)=1.042$\times$10$^{(1.719+0.762(B-V))}$, adapted from Tripicchio et al. (1999).\\
\\
\end{table*}

\section{The photometry of active solar-type stars}
\label{sec4}

We have performed a series of simulations under different conditions and
obtained final values of ``observed'' (predicted) photometry for a main
sequence star. This star has an effective temperature corresponding to that of
the quiet region, and we have assumed  different filling factors for the spots
(in principle, the photometry is not affected by the presence of faculae). We
have used different spots: one of them corresponds to the average of the actual
sunspot. The other one is the central part of it --the nucleus of the umbra--,
with lower temperature (4870 and 4575 K, respectively).
We have also carried out the computations for spots at 4000 and 3700 K.
 The model was applied
to stars with the following initial temperatures: 6344, 5655,  4988 and 4500 K,
corresponding to (B--V)=0.500, 0.679, 0.900 and 1.100, respectively.
Table 2 lists the initial model parameters (magnitudes, colors, radii, effective temperatures).  
The initial values of potassium and sodium equivalent widths
adopted for them are shown in Table 3.
Lithium equivalent widths and abundances are listed in Table 4a and Table 4b, in the case 
of the solar-like case and the Pleiades, respectively.

Figure 1a shows the variation of the visual magnitude ($\Delta$M$_V$, defined as
the difference between the magnitude of a star only covered by quiet regions 
and the observed magnitude; always positive) against the filling factor of the
spots, F$_{\rm spot}$. 
The solid line shows the results for a spot at 4870 K,
the dotted line represents the calculation for a spot at 4575 K, 
whereas the short and long  dashed lines represent the values for spots with 
4000 and 3700 K, respectively.  Although
the luminosity can decrease by a factor 10 in the most extreme case, a more
realistic situation can be considered in the range 0.0$\le$F$_{\rm
spot}$$\le$0.5, which, depending on these particular four spots, corresponds to
maximum variations of $\Delta$V $\sim$ 0.70 -- 0.60 mag. These values have been
observed in several active binary systems such as CF Tuc, BD$+$25$^\circ$161,
AY Cet, CC Eri, UX Ari, etc (e.g., Strassmeier et al. 1993; Strassmeier 1995,
and references therein). 

Figure 1b displays the dependence of the observed color index (B--V) on the
spot filling factor. It is seen that the color can change considerably, a
phenomenon which could have important consequences when comparing stars having
a high level of activity with inactive stars (i.e., rapid versus slow rotators)
at the same observed color, because  the color of the quiet photosphere is
bluer than the observed color. 
Note the different behavior for different T$_{\rm quiet}$,
due to the fact that the V filter is centered around $\sim$5500 \AA,
which corresponds to a black-body temperature between the spots
at 4870 and 4575 K, whereas the B filter is centered at  $\sim$4300 \AA.
Therefore, the relative contributions at the filters B and V are very different
depending on the spot and quiet photosphere temperatures.
Under realistic conditions, this change can be
up to $\Delta$(B--V) $\sim$ 0.1. Since there is a relationship between the color
indices and the stellar mass,  the mass estimated for the more active star
would be lower than its actual value in the case of MS stars. Using the
presence of photometric variations due to the modulation with phase, as the
star rotates, can help to avoid this bias, although the lack of modulation does
not guarantee the absence of this bias, since the spots can be distributed in a
homogeneus way on the stellar disk, producing null or very small photometric
variations. In fact, studies made with tomographic spectroscopy or Doppler
imaging (Strassmeier 1995; Vogt \& Hatzes 1995) and theoretical works
(Sch\"ussler 1995), show that the spots in very active systems tend to be
concentrated in the poles, staying there for a long period of time (time scale of
years), and producing minimum photometric variations.

\setcounter{table}{3}

\begin{table*}
\caption[ ]{{\bf a} Lithium equivalent widths and abundances  for  a solar-like case.}
\begin{tabular}{ccccccccc}
\hline 
Model  & 
\multicolumn{2}{c}{Temperature (K)}             &\,&
A(Li)  &   W(Li)  &\,&
A(Li)  &   W(Li)  \\
\cline{2-3}  \cline{5-6} \cline{8-9} 
     &   Quiet & Spot &&     Quiet    &    Quiet    &&    Spot     &   Spot     \\
(1)  &   (2)   &  (3) &&     (4)      &    (5)      &&    (6)      &    (7)     \\
\hline           	          
 1a  &   6344  & 4870 && 1.0 &  1.010$^{(1)}$  && 0.85 &  20.5$^{(2)}$ \\
 1b  &   6344  & 4575 && 1.0 &  1.010$^{(1)}$  && 0.87 &  49.1$^{(2)}$ \\
 1c  &   6344  & 4000 && 1.0 &  1.010$^{(1)}$  && 1.00 & 173.0$^{(1)}$ \\  
 1d  &   6344  & 3700 && 1.0 &  1.010$^{(1)}$  && 1.00 & 249.8$^{(3)}$ \\  
       	                                                      	    	
 2a  &   5655  & 4870 && 1.0 &  3.977$^{(2)}$  && 0.85 &  20.5$^{(2)}$ \\
 2b  &   5655  & 4575 && 1.0 &  3.977$^{(2)}$  && 0.87 &  49.1$^{(2)}$ \\
 2c  &   5655  & 4000 && 1.0 &  3.977$^{(2)}$  && 1.00 & 173.0$^{(1)}$ \\
 2d  &   5655  & 3700 && 1.0 &  3.977$^{(2)}$  && 1.00 & 249.8$^{(3)}$ \\
	                         	                             	  
 3a  &   4988  & 4870 && 1.0 &   20.4$^{(1)}$  && 0.85 &  20.5$^{(2)}$ \\
 3b  &   4988  & 4575 && 1.0 &   20.4$^{(1)}$  && 0.87 &  49.1$^{(2)}$ \\
 3c  &   4988  & 4000 && 1.0 &   20.4$^{(1)}$  && 1.00 & 173.0$^{(1)}$ \\
 3d  &   4988  & 3700 && 1.0 &   20.4$^{(1)}$  && 1.00 & 249.8$^{(3)}$ \\
      	                                                      	    	
 4c  &   4500  & 4000 && 1.0 &   77.0$^{(1)}$  && 1.00 & 173.0$^{(1)}$ \\
 3d  &   4500  & 3700 && 1.0 &   77.0$^{(1)}$  && 1.00 & 249.8$^{(3)}$ \\
\hline
\end{tabular}
$\,$\\
$^{(1)}$ From Soderblom et al. (1993) curves of growth, using A(Li)=1.00.\\
$^{(2)}$ Measured. See Table 1.\\
$^{(3)}$ From Pavlenko et al. (1996) curves of growth, using A(Li)=1.00.\\
\\
\end{table*}

\setcounter{table}{3}

\begin{table*}
\caption[ ]{{\bf b} Lithium equivalent widths and abundances  for  Pleiades stars.}
\begin{tabular}{cccccccc}
\hline 
Model  & 
\multicolumn{2}{c}{Temperature (K)}             &\,&
A(Li)  &\,&   \multicolumn{2}{c}{W(Li)}  \\
\cline{2-3}  \cline{5-5} \cline{7-8} 
     &   Quiet & Spot &&     Quiet    &&   Quiet &   Spot     \\
(1)  &   (2)   &  (3) &&     (4)      &&   (5)   &    (6)     \\
\hline           	                		  
 1a  &   6344  & 4870 && 3.02$^{(1)}$ &&  80.0  &  313$^{(1)}$ \\
 1b  &   6344  & 4575 && 3.02$^{(1)}$ &&  80.0  &  383$^{(1)}$ \\
 1c  &   6344  & 4000 && 3.02$^{(1)}$ &&  80.0  &  605$^{(1)}$ \\  
 1d  &   6344  & 3700 && 3.02$^{(1)}$ &&  80.0  &  993$^{(2)}$ \\  
       	                                          	   
 2a  &   5655  & 4870 && 2.80$^{(1)}$ && 140.0  &  284$^{(1)}$ \\
 2b  &   5655  & 4575 && 2.80$^{(1)}$ && 140.0  &  345$^{(1)}$ \\
 2c  &   5655  & 4000 && 2.80$^{(1)}$ && 140.0  &  510$^{(1)}$ \\
 2d  &   5655  & 3700 && 2.80$^{(1)}$ && 140.0  &  632$^{(2)}$ \\
	                                  	              	   
 3a  &   4988  & 4870 && 1.70$^{(1)}$ &&  84.0  &  107$^{(1)}$ \\
 3b  &   4988  & 4575 && 1.70$^{(1)}$ &&  84.0  &  177$^{(1)}$ \\
 3c  &   4988  & 4000 && 1.70$^{(1)}$ &&  84.0  &  276$^{(1)}$ \\
 3d  &   4988  & 3700 && 1.70$^{(1)}$ &&  84.0  &  382$^{(2)}$ \\
      	                                          	   
 4c  &   4500  & 4000 && 0.64$^{(1)}$ &&  37.0  &  112$^{(1)}$ \\
 3d  &   4500  & 3700 && 0.64$^{(1)}$ &&  37.0  &  157$^{(2)}$ \\
\hline
\end{tabular}
$\,$\\
$^{(1)}$ From Soderblom et al. (1993) curves of growth, using W(Li) --quiet-- from Figure 6a.\\
$^{(2)}$ From Pavlenko et al. (1996) curves of growth, using W(Li) --quiet-- from Figure 6a.\\
\\
\end{table*}

We have performed several comparisons between the observed data and the
predictions of this model on the effect of the presence of surface
inhomogeneities on the photometry and the equivalent widths. A first check is
provided by the Pleiades open cluster by looking at the location of the active
stars in the Color-Magnitude diagram (CMD), where the photometric properties of
the late-type stars change due to the presence of surface features with
different temperatures. Figure 2 shows the visual absolute magnitude against
the dereddened color index (B--V) for  the four  cases presented in this study. The
behavior of the observed photometry was computed using a quiet photosphere
with four different color indices (0.500, 0.679,  0.900 and 1.100) corresponding to the
 effective temperatures listed in Table 2. The main sequence  appears as a thin solid
line, whereas the locus  for twin stars (MS--0.75$^{\rm mag}$) is represented as a dotted line
parallel to it. Real data for stars belonging to the Pleiades
(Soderblom et al. 1993) are included as open circles and the size increases
with increasing stellar activity, measured using the Ca\,{\sc ii} IR triplet at
8542 \AA\ (R$_{\lambda=8542}$). Those Pleiades stars without measured 
calcium appear as plus symbols.
Solid diamonds in the figure represent model predictions for 
F$_{\rm spot}$=0.50.
 Our model shows that the presence of spots
shifts the location of these stars almost in parallel with the MS, except when the 
difference between the quiet atmosphere and the stellar spot becomes very large. In this last
 case, when  the spots are  at
temperatures much lower than the photospheric value, large
increments of the observed magnitude  with respect to the MS at the same observed color
could appear, with almost no change in the color.
 However, the lack of stars in this region suggests that either
such differences are not possible (i.e., T$_{\rm quiet}$=6344 K 
and T$_{\rm spot}$=3700 K) or the filling factor must be very small in this case.
 Therefore, for a given observed color index, the active stars
would tend to be located in the lower part of the MS (the less luminous
region), even below the MS of single and inactive stars. In the most extreme
case, they could be found below the Zero Age Main Sequence (ZAMS). However, the
comparison with the Pleiades shows that this phenomenon is not observed.
Several factors can explain this situation: first, the physical size of the
cluster can introduce a scatter in the CMD equivalent to several tenths of
magnitude, which can hide the effect of the activity on the photometry (i.e.
Barrado y Navascu\'es \& Stauffer 1996). On the other hand, an important
fraction of the stars are in fact binaries and each component can have a
different degree of activity, a case not included in the assumptions made in
our simple model. 
However, the effect of stellar spot on the observed photometry of star
suggests that it might be influencing on  the Hipparcos problem
 (i.e., the fact that the locus of very well known,
nearby open clusters whose distance have been measured by the Hipparcos satellite.
See Pinsonneault et al. 2000; Robichon et al. 2000, and references therein), 
in the sense that two open clusters of similar ages but very different 
rotational velocity distributions might have a shift in the locus of their
main sequence when compared with each other.
 In summary, Figure 2 shows that the  photometry obtained
using our model is compatible with the observational data and their
 uncertainties for situations showing not large differences between 
 T$_{\rm quiet}$ and T$_{\rm spot}$ and realistic values of F$_{\rm spot}$.

\begin{figure}
\resizebox{8.7cm}{!}{\includegraphics{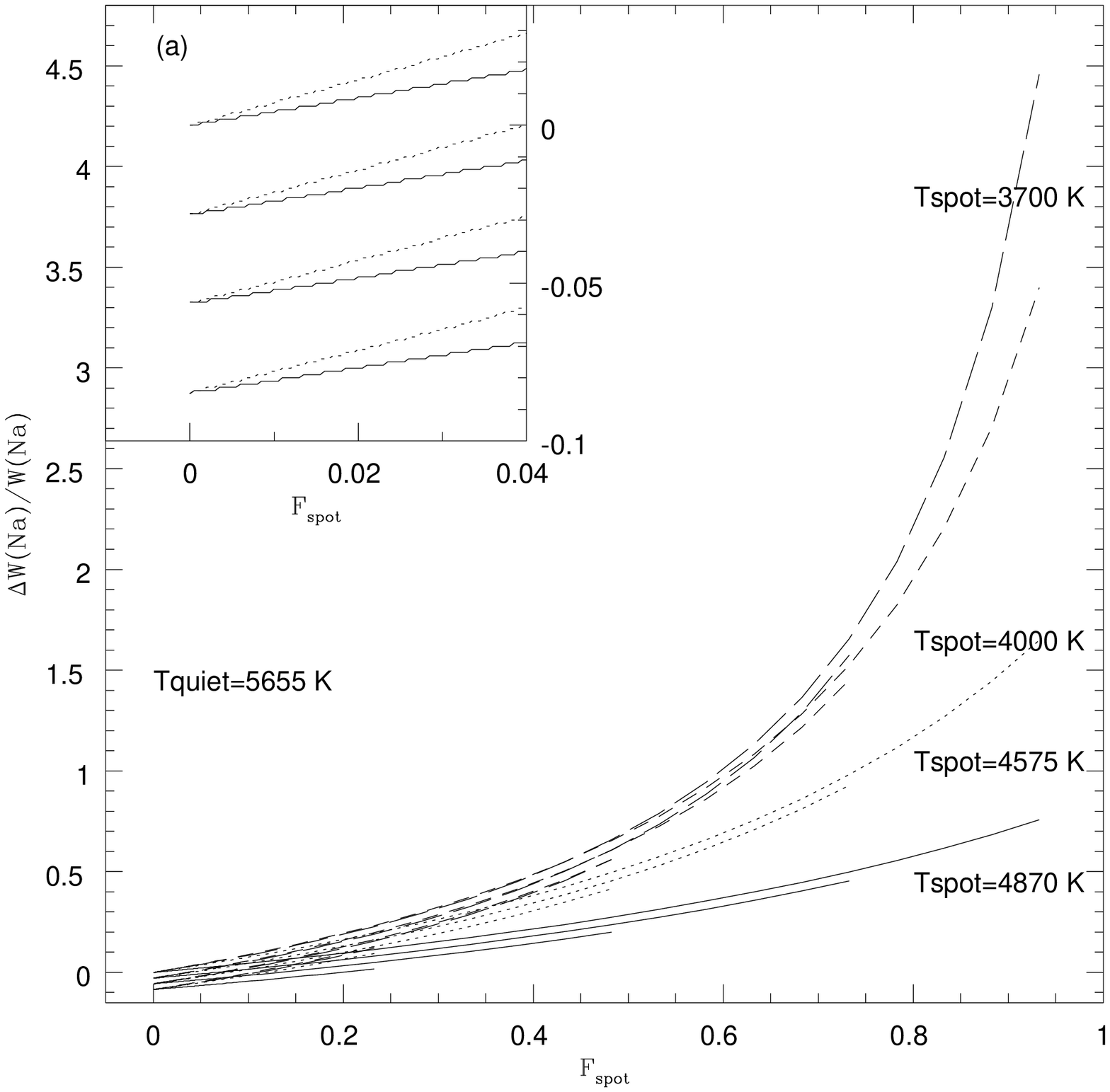}}
\resizebox{8.7cm}{!}{\includegraphics{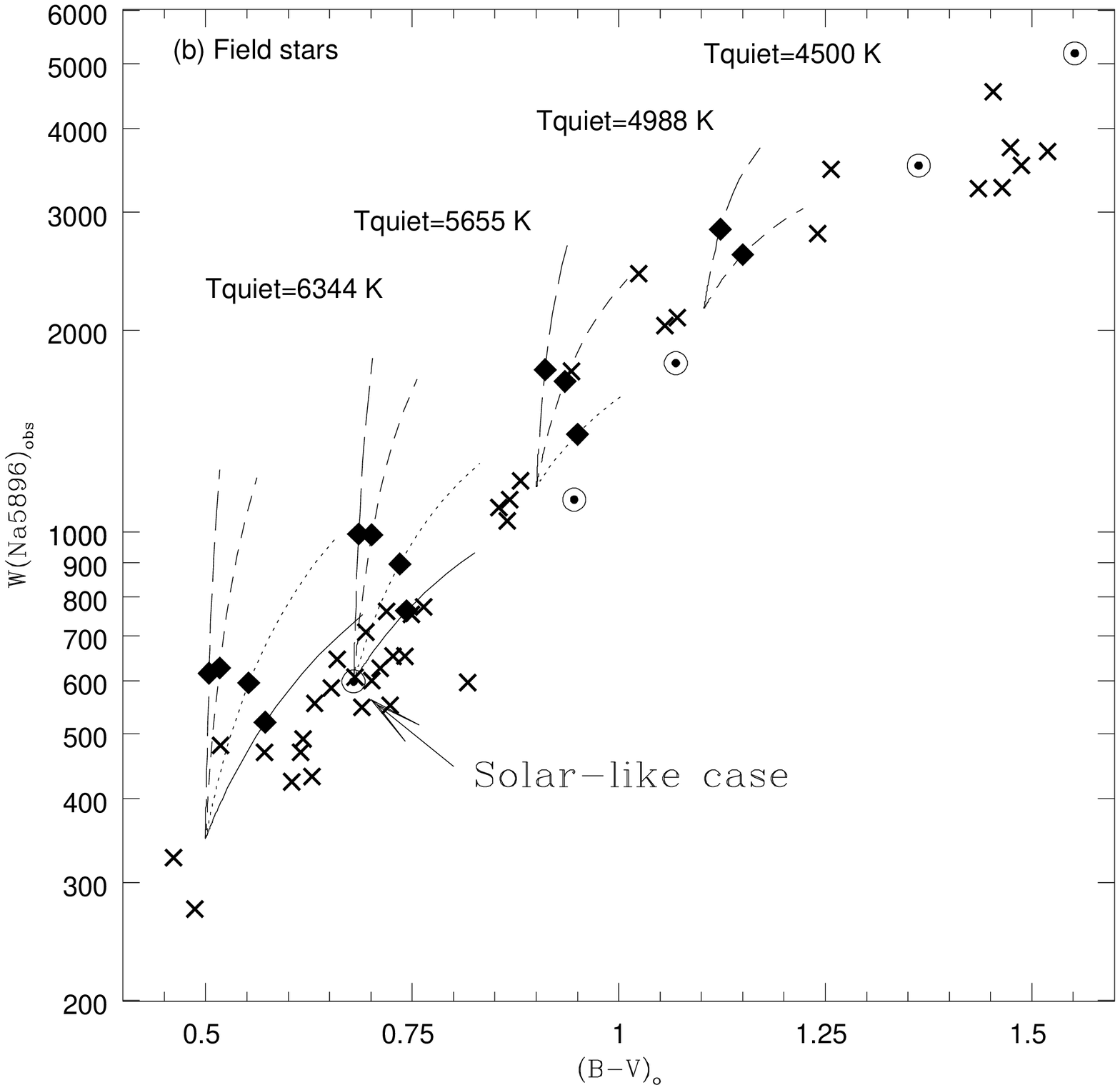}}
\caption[]{\label{fig4}
{\sl (a)} Variations of the sodium equivalent width against 
the spot filling factor.
 Symbols as in Figure 3a.  
{\sl (b)}  Variations of the sodium equivalent width against
(B--V) color index. Field dwarfs from Tripicchio et al. (1997) 
are displayed as crosses. 
}
\end{figure}

\section{The effect of stellar  spots and faculae on the alkali}
\label{sec5}

\subsection{Potassium (K\,{\sc i} 7699 \AA)}
\label{sec5.1}

The direct comparison between the W(K\,{\sc i} 7699 \AA) corresponding to
different regions of the solar photosphere indicates that the variations due to
the presence of spots have to be small, except for large values of the spot
filling factor. Figure 3a shows the relative variation of the observed W(K)
against F$_{\rm spot}$ for the solar-like case.
 In the most extreme cases, when F$_{\rm spot}\sim 1$,
changes close to a 300\% can be reached. However, the more realistic values
shown in the figure indicate changes between 0\% and 50\%. In any case, this
kind of variation due to spots should be easily observed, as we show below.

Figure 3a includes several sets of curves: the dotted lines represent the
results obtained using a spot at 4575 K and different values of the faculae 
filling factor. From top to bottom, F$_{\rm faculae}$=0.00, 0.25, 0.50 and 0.75.
The solid lines correspond to the computations for T$_{\rm spot}$=4870 K.
Other sets, computed with  T$_{\rm spot}$=4000 K and T$_{\rm spot}$=3700 K,
are shown as short and long dashed lines, respectively.
A detail of the graph appears in the box (we only display the first two cases).
Essentially, the values of F$_{\rm faculae}$ do not affect the observed value
of W(K), as a consequence of the small difference between the measured 
equivalent widths in
the quiet region and the faculae.

In an analogous way to the previous figure, it is possible to compare the
variations of the W(K) with (B--V) --the values that would be observed, in
both cases. There is a quasi-linear relation between both quantities.
However, a more interesting comparison is made in Figure 3b, where we have
included real data corresponding to the Pleiades cluster (Soderblom et al.
1993). The size of the symbols, as in Figure 2, increases with increasing
stellar activity (measured using R$_{\lambda=8542}$). As noted by Soderblom et
al. (1993), in the range 0.50$\le$(B--V)$\le$0.80 there is a clear trend
between activity and the potassium equivalent width for a given value of 
(B--V): the larger the activity, the larger W(K). The reproduction of this 
behavior has been attempted previously by Stuik et al. (1997), using a purely
theoretical model under NLTE conditions. However, their models were not able
to match the Pleiades observations (see their Figure 8). 

In order to verify if our empirical approximation would be able to match the 
behavior observed in the Pleiades, we proceeded as described in Section 3.
We computed the results using four
different situations: a star having a color (B--V)=0.500, another star having a
color corresponding to that one of the quiet region surrounding the sunspot
--(B--V)=0.679--, a third star with (B--V)=0.900,
 and another with (B--V)=1.100,
 the same values used for
the simulations shown in Figure 2. We used spots at four different
temperatures (4870,  4575, 4000 and 3700  K), indicated with
solid, dotted, short dashed and long dashed lines.  The solar symbols
(dots plus circles) indicate the location of the quiet sun and the data for the
solar spots. 
A comparison with field dwarfs, from Tripicchio et al. (1999) are displayed in
Figure 3c. In this case, field stars are shown as crosses. 
Unfortunately, information regarding the activity
of these stars is available for only a handful of them.

As  can be concluded from Figure 3b and Figure 3c, the behavior of W(K) is  well 
reproduced despite the simplicity of the procedure. It explains why active
stars tend to have larger equivalent widths.
 The model also explains why there is not a sharp
distinction between the location of inactive and active stars: it is predicted
that there are locations where it is possible to find stars having high and
intermediate activity, or intermediate and low activity. This is because stars
having different masses are compared: since the activity of the more massive
stars modifies the observed photometry to a greater extent, both stars (the
very active and the less active) have similar apparent colors. Note, however,
that realistic values of the spot filling factor (F$_{\rm spot}$$\le$0.50;
indicated by solid diamonds in Fig. 3b) cannot reproduce those stars having the
largest values of the W(K\,{\sc i} 7699) for a given color.
On the other hand, the figures should be interpreted with some caution, since
the model is rather sensitive to the initial conditions (i.e., the initial
effective temperature of the spot and the quiet photosphere and the
potassium equivalent width of the spot).

\begin{figure}
\resizebox{7.cm}{!}{\includegraphics{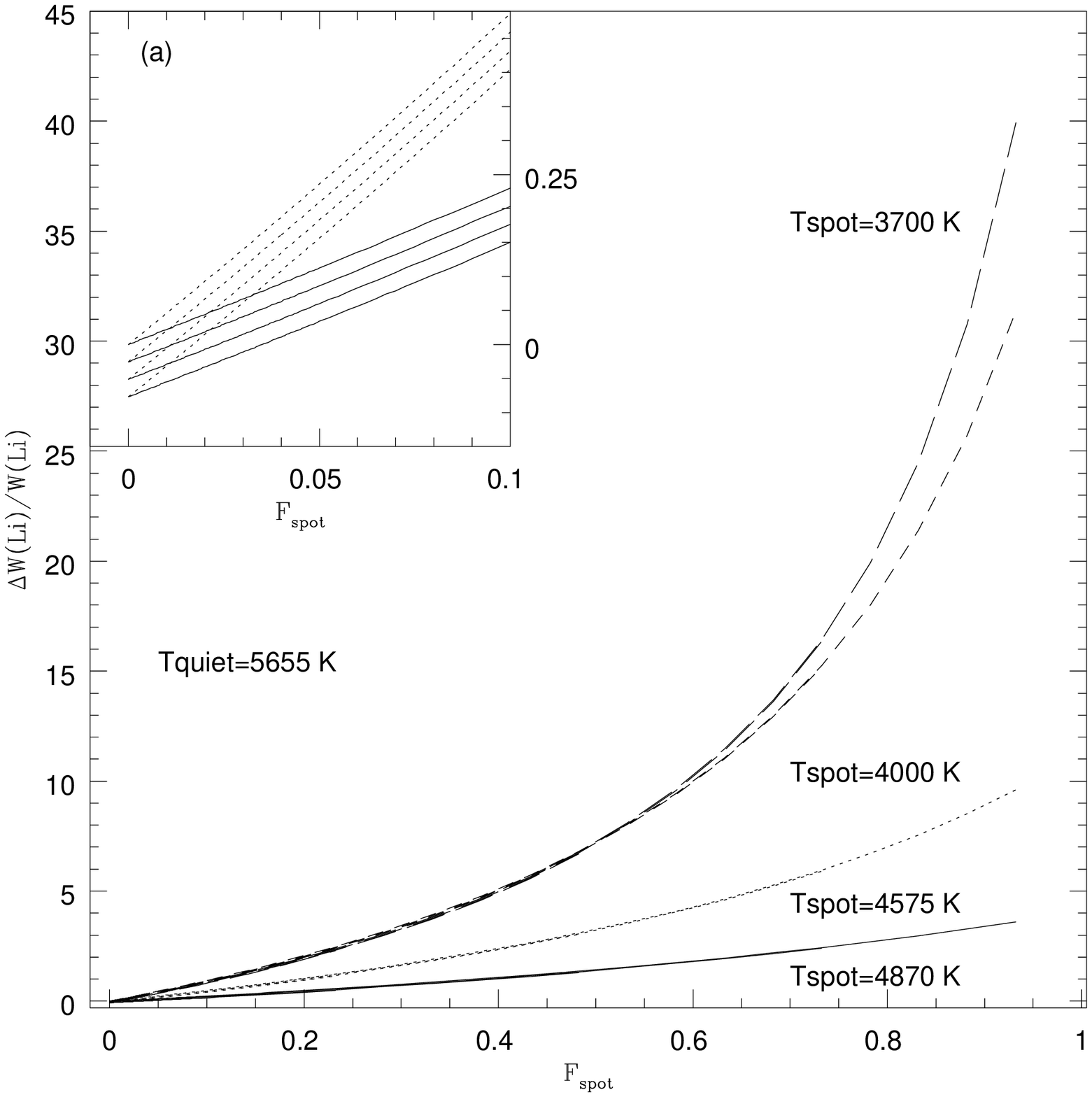}}
\resizebox{7.cm}{!}{\includegraphics{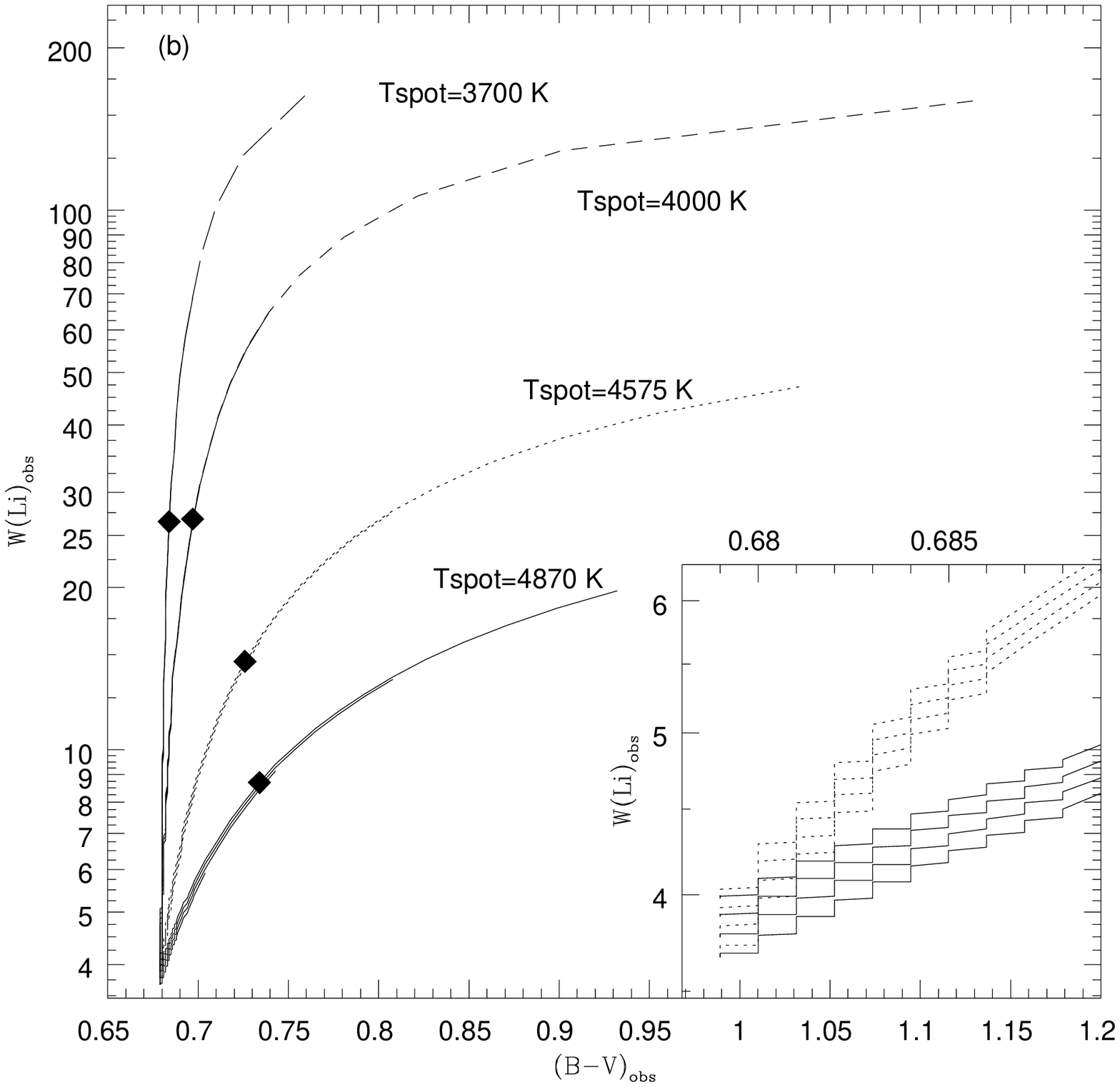}}
\resizebox{7.cm}{!}{\includegraphics{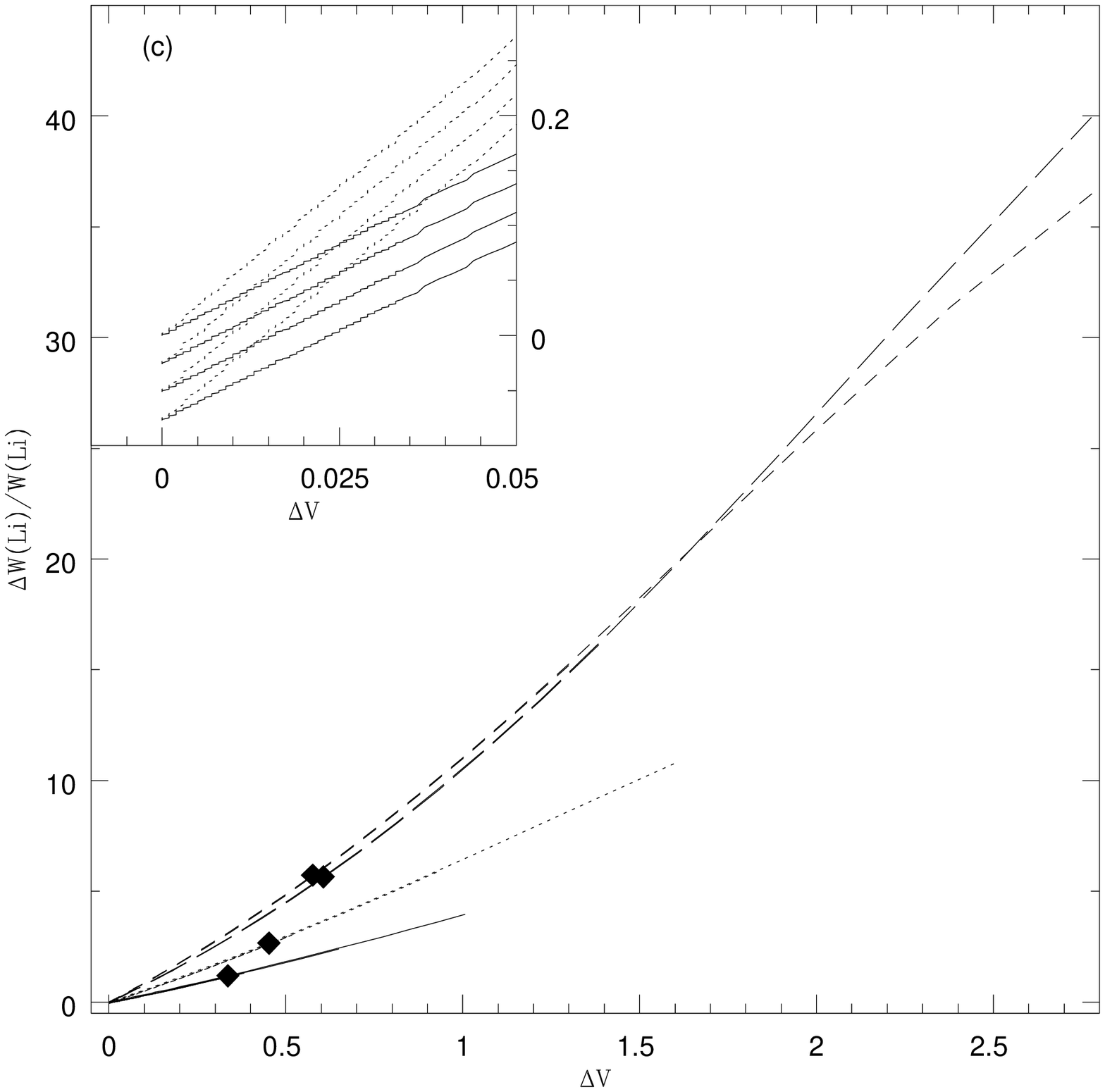}}
\caption[]{\label{fig5} 
{\sl (a)} Variations of the lithium doublet equivalent width against 
the spot filling factor.
{\sl (b)} Lithium equivalent width versus  (B--V) color index.   
{\sl (c)} Variations of the lithium doublet equivalent width against 
 photometric variations.   A detail of these  graphs
are displayed in the small boxes. Symbols as in Figure 3a.}
\end{figure}

\subsection{Sodium (Na\,{\sc i} 5896 \AA)}
\label{sec5.2}

As it is shown in Table 1, the W(Na) is much  larger in a sunspot
than in the quiet region. Therefore, the observed value of the equivalent width
 changes in a
much larger amount than in the case of  potassium for the same filling factor
of the spot, although the relative variation is similar for both K\,{\sc i} and
Na\,{\sc i} (with respect to their respective equivalent widths). Figures 3a
and 4a illustrate this fact. In addition, the model predicts that the observed
value for Na\,{\sc i} could be smaller than the equivalent width corresponding
to the star when it is inactive. This unlikely situation would appear if an
important fraction of the surface were covered by active regions (faculae) but
having very small spot coverage. In any case, the important result coming from
this figure is that it is possible to have a situation in which a high degree
of activity corresponds with no variations of the W(Na). This would happen if
the active regions were distributed homogeneously, or  for certain combinations
of the  values of F$_{\rm spot}$ and F$_{\rm faculae}$.

Figure 4b plots predicted W(Na) against (B--V).  We have included in the 
figure, as crosses,  data corresponding to field dwarfs (Tripicchio et al. 1997).
Note that the field stars could have different abundances of sodium.
However, assuming a similar abundance, 
our model is able, in principle, to explain the scatter of the observed
W(Na).
Although, to our knowledge, there are no
studies about the behavior of this doublet in late-type stars for young
clusters, we would expect in principle a similar behavior to the case of
K\,{\sc i} 7699 \AA\ (i.e., for a given color, the larger the activity, the
larger the equivalent width). However, there is a significant difference
between Na\,{\sc i} 5896 and K\,{\sc i} 7699 \AA\ lines: the significant
difference of equivalent widths  between faculae and quiet regions for Na\,{\sc
i}, which is not observed for K\,{\sc i}. So, the scatter displayed in Figure
3b (potassium  equivalent width against color index) might be different in a
similar plot for sodium. Therefore, a sample of coeval stars of different
colors might not show a clear relationship between activity and W(Na), since a
star can have a large activity and, simultaneously, change minimally 
the equivalent width of Na\,{\sc i} 5896 \AA\ (if a large fraction of the
stellar surface is covered with active regions). This is a very
interesting tool to verify the effect of spots and faculae on the equivalent widths of
alkali.

\begin{figure}
\resizebox{8.7cm}{!}{\includegraphics{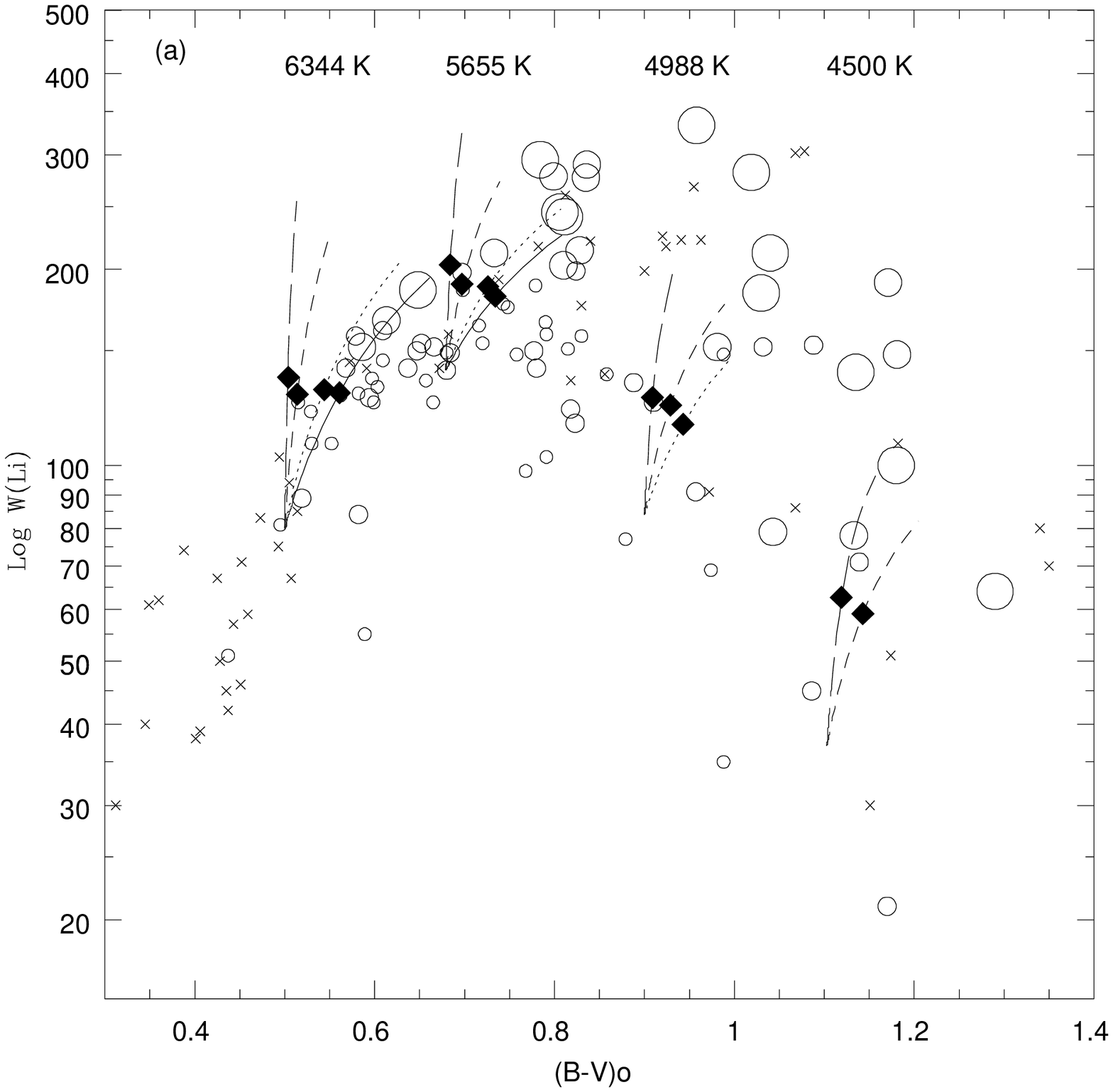}}
\resizebox{8.7cm}{!}{\includegraphics{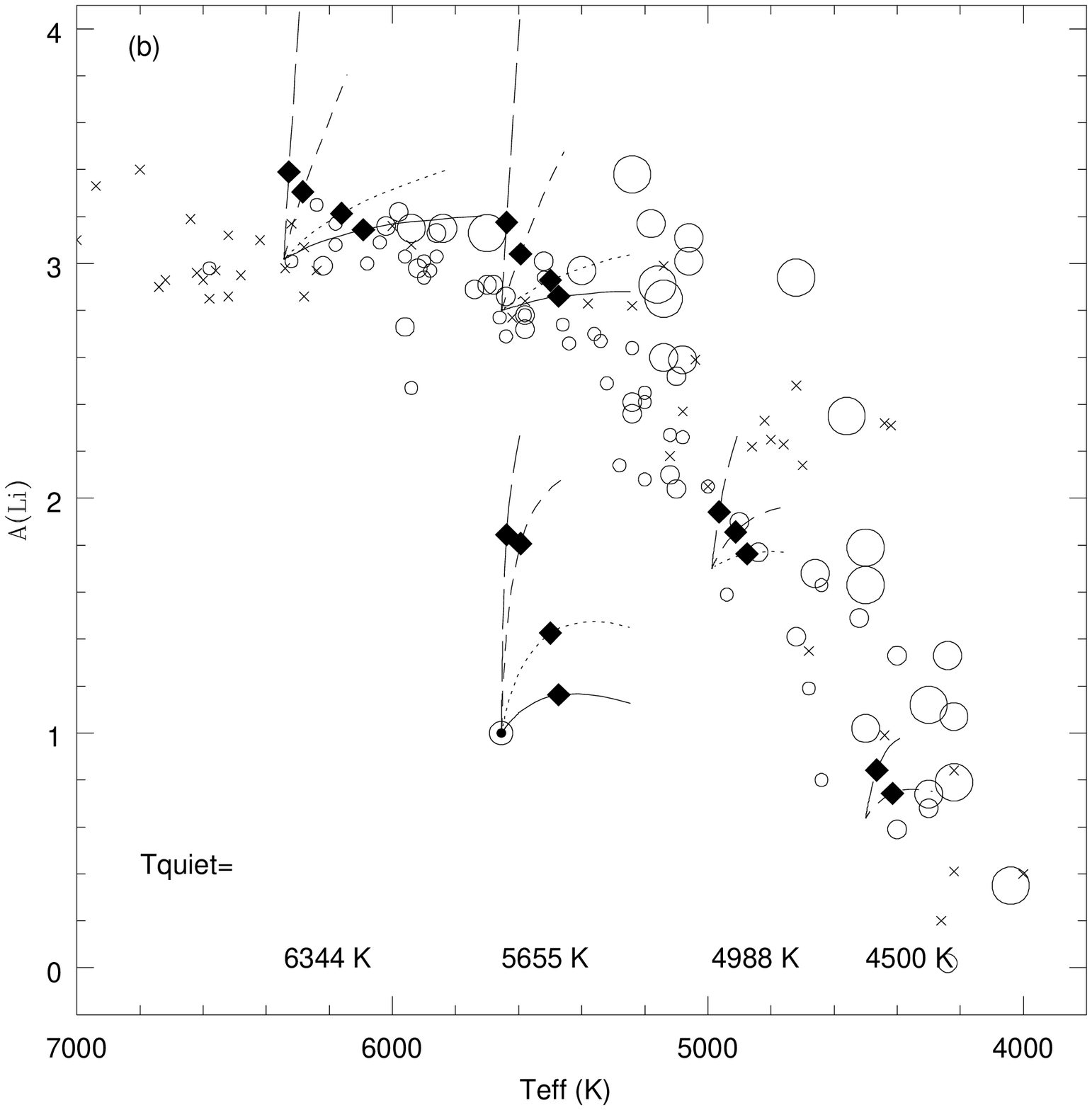}}
\caption[]{\label{fig6} Comparison with Pleiades data:
{\sl (a)} \ Lithium 
equivalent width against color. 
{\sl (b)} \  Lithium abundance against effective 
temperature. Symbols as in Figure 2. We have also
included the case for the Sun.}
\end{figure}

\begin{figure*}
\resizebox{18.cm}{!}{\includegraphics{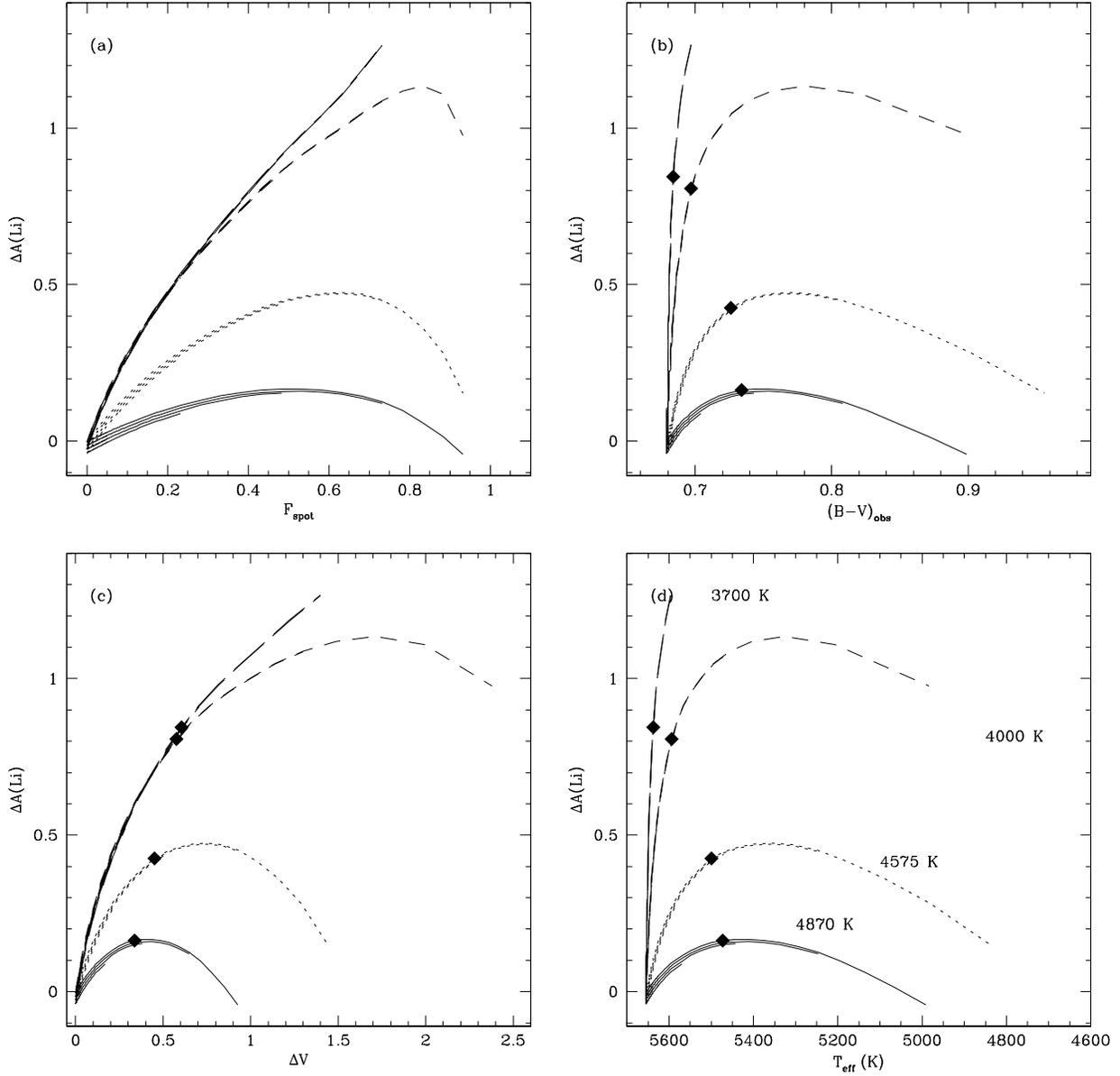}}
\caption[]{\label{fig7} Behavior of the differences of lithium abundances
(A(Li)$_{\rm model}$--A(Li)$_{\rm initial}$)
against spot filling factor {\sl (a)}, 
(B--V) index {\sl (b)}, 
photometric variations  {\sl (c)},
 and effective temperature {\sl (d).}
Symbols as in Figure 3a. In all cases A(Li)$_{\rm initial}$=1.00 dex. 
}
\end{figure*}

\subsection{Lithium (Li\,{\sc i} 6708 \AA)}
\label{sec5.3}

\subsubsection{The behavior of the W(Li)}
\label{sec5.3.1}

The study of the variation of the lithium equivalent width due to the presence
of surface inhomogeneities (spots and faculae) can provide essential hints to
understand the abundances derived for stars with a high level of activity, such
as chromospherically active binary systems and rapid rotators belonging to
young clusters.

Our model predicts a large variation in W(Li), even close to 3500\%, due to the
presence of spots. This can be clearly seen in Figure 5a. This kind of 
variation has already been predicted by Giampapa (1984). The behavior of the
lithium line is very sensitive to the temperature of the spot.

As  for Na\,{\sc i} 5896 \AA, the presence of faculae is  an
important factor that can alter the observed equivalent widths
 to a significant degree, even
reducing it when comparing with the characteristic value of a quiet region.
Figure 5a shows that a solar-like star could have null variations of W(Li) under a
wide variety of situations, all of them  realistic. This occurs when the
filling factor of the spots is small  and a significant part of the
photosphere is covered with faculae.

Figure 5b shows the extraordinary change of the W(Li) as the observed (B--V)
color increases in the solar-like case,
 as a consequence of the presence of active regions. The
dependence on the spot temperature is evident: the smaller the
temperature, the larger the observed W(Li) and color.

Pallavicini et al. (1993), in a quasi-simultaneous study involving photometry
and spectroscopy (measuring Li\,{\sc i} 6708 and Ca\,{\sc i} 6718 \AA) of 4 
active stars, did not observe any change in the equivalent width of the Li
line, even when these objects presented important photometric variations (up to
0.10 mag). Our model provides an explanation for this apparent puzzle. Although
the values of F$_{\rm spot}$ which correspond to $\Delta$V=0.10 mag could
produce important increments in W(Li), taking into account the
faculae in the simulation changes the scenario completely. The region covered
by faculae has a very reduced value of the W(Li) and tends to dilute the line
produced by the quiet region and by the spot. 
Moreover, the presence of faculae is not detected in the photometric
modulation. It is reasonable to assume that the fraction of the stellar disk
covered by faculae has to be at least as large as the fraction covered by
spots. The final consequence is that a star having high or moderate activity
can experience important photometric variations together with reduced
variations in the Li equivalent width (either positive or negative), as Figure
5c demonstrates, where we represent the relative
variation of the equivalent width of lithium against the photometric variation.

\subsubsection{A second comparison with the Pleiades}
\label{sec5.3.2}

An additional check to our model can be performed by comparing it, again, with
Pleiades data. Figure 6 shows actual Li data (Soderblom et al. 1993), where
the size of the symbols increases with increasing activity. Simulations with
different photospheric temperatures are displayed, computed with spots with
four temperatures (4870, 4575,  4000 and 3700 K; solid,  and dotted, short and long dashed 
lines, respectively).
 As with potassium, the spread on the observed lithium
equivalent  widths in the Pleiades can be reproduced (Figure 6a).
However, in the lithium case, only part of the spread could be due to the effect
of the stellar inhomogeneities, since our curves, even for very large filling factors, 
do not cover the whole area where the Pleiades stars are located. 
However, the
simulations can explain the trend with stellar activity (for a given color): 
active stars have a larger value of W(Li). Therefore, part of the spread in
the equivalent widths could be due to the simultaneous effect of the surface
inhomogeneities on the observed equivalent width and color. Some of the cases
would require, however, a (non--realistic) very large filling factor or a large
difference between the temperature of the quiet photosphere and the spot. 

The A(Li)-T$_{\rm eff}$ plane (Figure 6b, where A(Li)$=\log ({\rm Li/H})+12$)
provides complementary information. Predicted equivalent widths were translated into Li
abundances using the curves of growth by Soderblom et al. (1993). 
In the case of temperatures below 4000 K (the lowest temperature of these
curves of growth), we used those published by Pavlenko et al. (1996).
Predicted
abundances tend to move the observed values to lower effective temperatures for
the most active stars, therefore reducing the spread in the abundance and the
rotation-activity relation. 
 The lithium abundances for active stars appear then to be correct but
their location should be shifted toward higher temperatures in the 
A(Li)-T$_{\rm eff}$ plane due to the effect of the spots on the colors. We must note,
however, that only very large activity levels would change the morphology of
the T$_{\rm eff}$-A(Li) plane. Even spot filling factors close to  F$_{\rm
spot}$=0.50 move the location of a star in this diagram by $\sim$180 K only
towards cooler temperatures.
On the other hand, as discussed before, our simulations do not explain the high 
lithium abundances in the range 5200--4600 K. Therefore, it seems that 
these abundances are real and that an important fraction the lithium  spread 
which is present in the figure is real, and that the rotation is a key 
factor in the evolution of the lithium abundance in the late spectral type stars.

\begin{figure}
\resizebox{8.7cm}{!}{\includegraphics{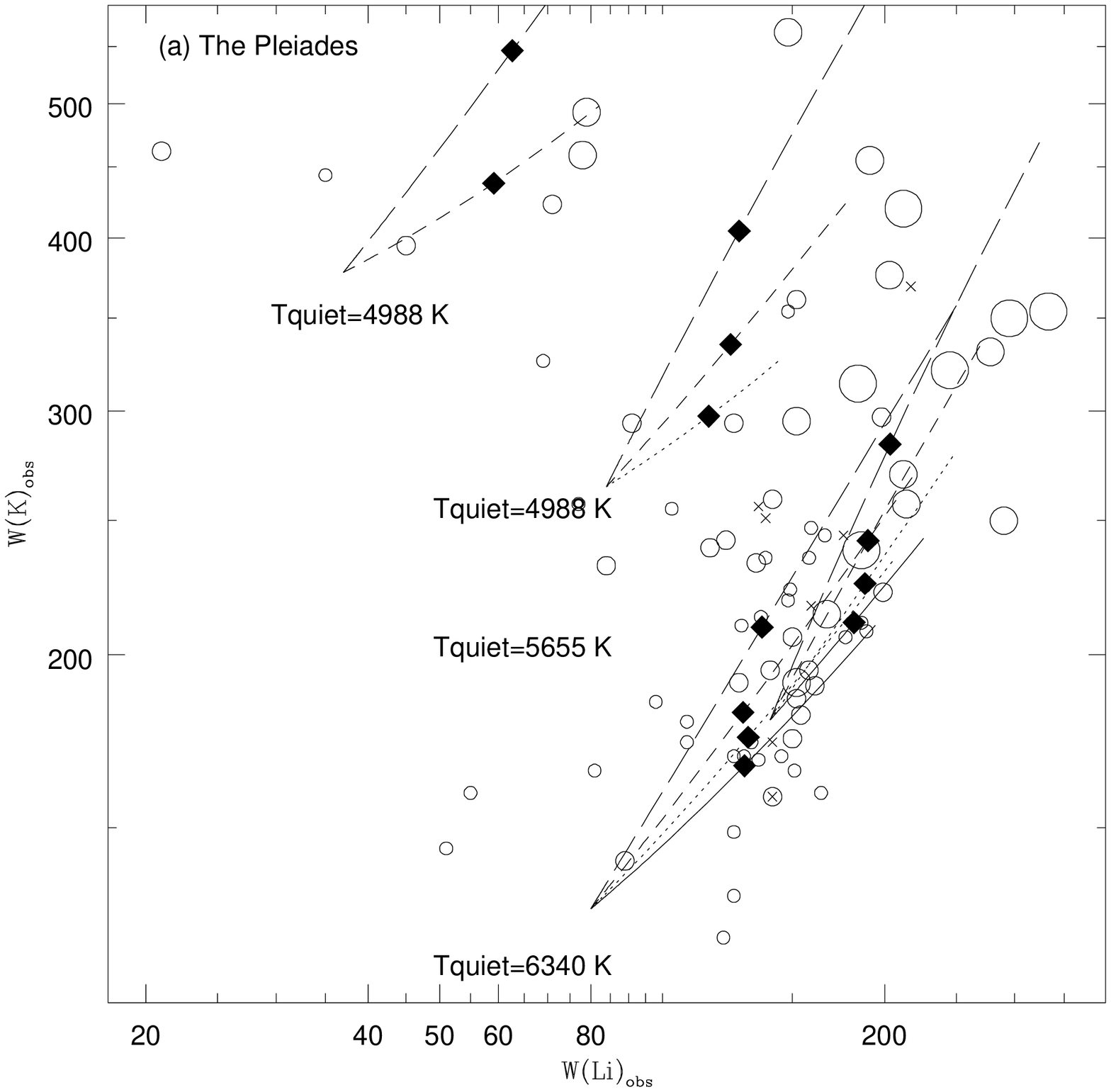}}
\resizebox{8.7cm}{!}{\includegraphics{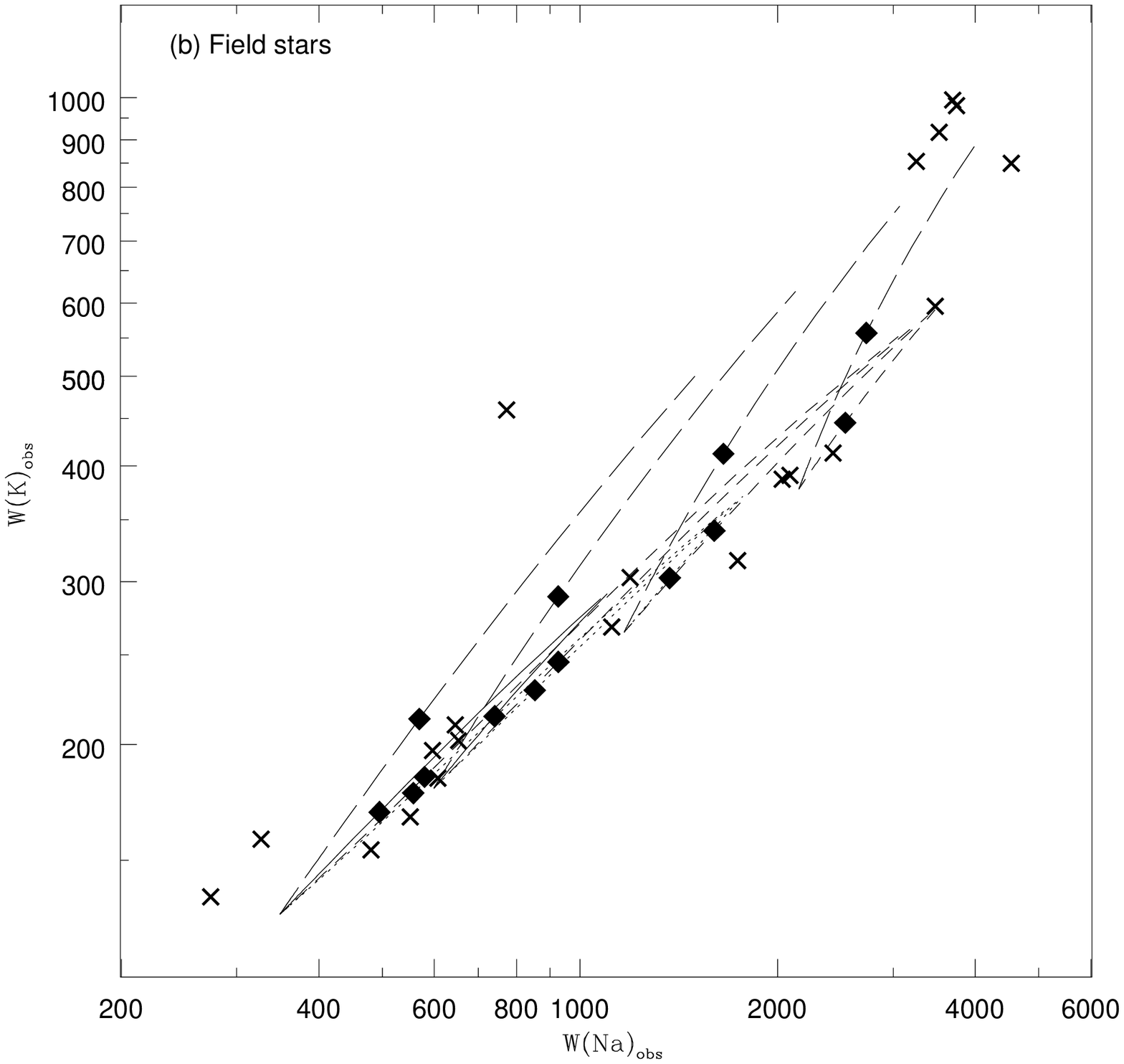}}
\caption[]{\label{fig8}
{\sl a}
 Comparison between the relative variations of
potassium and lithium for the Pleiades. 
Symbols as in Figure 3b.
{\sl b}
 Comparison between the relative variations of
potassium and sodium for field stars. 
Symbols as in Figure 3c.
The last point in our calculations
corresponds to F$_{\rm spot}$=0.80.}
\end{figure}

\subsubsection{Uncertainties in the lithium abundance due to the presence of
active regions}
\label{sec5.3.3}

We have shown above (see Figure 5 and Figure 6)
 that the presence of inhomogeneities on the 
stellar surface could induce important variations in the measured Li equivalent
widths. It is extremely important to establish how these features affect the
accuracy of the derived lithium abundances. The presence of spots affects the
equivalent width in two different ways: the lithium equivalent width increases,
producing a pseudo-abundance larger than the real value. Simultaneously, the
color indices are reddened and the lower temperature favors values lower than
those corresponding to the quiet photosphere, inducing  a reduction in the
estimated abundance.  

Figure 7a shows the behavior of the abundance provided by the solar model against
the spot filling factor for a given initial abundance (A(Li)=1.00 in this
case). A temperature of 5655 K for the quiet photosphere, and temperatures of
4870 K (solid lines), 4575 K (dotted lines),  4000 K (short dashed lines),
and 3700 K (long dashed lines) for the spots were used.
Different filling factors for the faculae (from top to bottom, 0.00, 0.25, 0.50
and 0.75) were considered. It is clearly seen that the presence of
inhomogeneities could introduce important errors in the derived lithium
abundance. In this case, they span from $-0.05$ to $+$1.2 dex in the worst
cases. The decrease in the observed abundance appears when a large fraction of
the photosphere is covered by faculae and the filling factor of the spots is
small. It is more important when there is a large difference between the
spot and quiet photosphere temperatures.
These uncertainties are restricted, however, by realistic estimates of
the spot filling factor which would yield $\Delta$A(Li)$\sim$0.20 dex.

Figures 7b-d show the behavior of the lithium abundance against the observed
color (B--V), the increment in the visual magnitude, and the effective
temperature, respectively. These values were determined from (B--V) and the
Thorburn et al. (1993) temperature scale. The different observed parameters
can change by a considerable amount due to the presence of inhomogeneities.
However, the final uncertainties in the derived lithium abundances are limited
by the expected values of the filling factors.  

Doppler imaging techniques show that active stars have a tendency to
concentrate their spots in the poles (Vogt \& Hatzes 1995). Doppler imaging of
active Pleiades stars shows that they are, indeed, covered with huge polar
spots of presumably long life span (Stout-Batalha \& Vogt 1999). If this
controversial fact is true (see Byrne 1995 for a comprehensive discussion about
this subject), active stars would show little photometric variations due to
inhomogeneities and, simultaneously, the measured equivalent widths would be
affected. Therefore, these stars would be several tenths of magnitude fainter
than analogous inactive stars (same mass, chemical composition and age), and
could show different equivalent widths of the alkali for the same abundances
--in the case of lithium, W(Li)$_{\rm measured}$ could be larger--. However,
this situation is not observed when a comparison of the locations of active
and inactive stars in Color--Magnitude diagrams or W(Li)-(B--V) plane is
performed.

\subsection{Comparison between the alkali}
\label{sec5.4}

One of the aims of this work was to compare the behavior of the three alkali
elements under several circumstances, establishing the relative variations of
the observed equivalent widths in stars with different filling factors of spots
and faculae. Figure 8a shows the comparison between 
 potassium and lithium equivalent widths. Pleiades data are included.
 Symbols are
as in Figure 3b. Active stars tend to be in the top, right-hand  side
of the diagram. This behavior is qualitatively reproduced with our model.

 Figure 8b displays the 
the equivalent widths of sodium and potassium. Data corresponding to field
stars are shown as crosses. 
 The dependence of the trend
with the spot temperature seems to be very small. It would be very interesting 
to have similar data for members of a cluster, to determine the effect of the 
activity on a sample of stars with the same age.

\section{Conclusions}
\label{sec6}

A simple model has been constructed to study the effects of surface active
regions on the behavior of the Li\,{\sc i} 6708, Na\,{\sc i} 5896, and  K\,{\sc
i} 7699 \AA\ lines in the atmospheres of late-type stars. These effects
strongly depend on the actual distributions of surface inhomogeneities (spots
and faculae in particular) which, unfortunately, are not well known for those
objects. They could be distributed in irregular or homogeneous ways and the
changes in the alkali line equivalent widths will be affected by the filling
factors of the spots and faculae, temperatures of the different regions and the
element abundances. Additional observations providing independent estimates of
the filling factors, such as direct measurements of the magnetic field via
Zeeman splitting (Saar 1995) or empirical relations between the Rossby number
and the filling factors (Montesinos \& Jordan 1993), would help to determine
the actual values of the colors and equivalent widths and to constrain the
input model parameters.

Our simulations predict that active stars can suffer changes in their lithium
equivalent widths compared with quiescent objects. However, under normal 
circumstances (spot filling factors less than 0.5), these variations would not
induce significant changes in the derived abundances. A similar conclusion was
reached by Pallavicini et al. (1993) by studying the photometry and the
Li\,{\sc i} 6708 \AA\ line in several stars. Moreover, Pallavicini et al.
(1987, 1992), Randich et al. (1993, 1994), Fern\'andez-Figueroa et al. (1993),
and Barrado y Navascu\'es et al. (1997, 1998) have studied a large sample of
chromospherically active binaries in a time span of several years, and have
found no evidences of changes in the lithium equivalent widths that could be
related to the phase or to secular changes in the activity. 

The comparison between the model predictions and the observations suggests that
part of the spread observed in the equivalent widths of K and Li  lines, as
well as in the Li abundances, for the Pleiades could be due to the simultaneous
effect of surface inhomogeneities on the observed equivalent widths and
photometric colors. At this point it is important to obtain additional 
information on these lines for other young open clusters where a similar
dispersion in Li abundances has been found (eg. Alpha Persei cluster;
 Randich et al.
1998)  to check the model predictions in those cases.  Na\,{\sc i} D
measurements in the Pleiades and other clusters will provide complementary data
to witness the  behavior of the other two alkali elements.

\acknowledgements{
DBN acknowledges the support by the    
the {\it ``Real Colegio Complutense''} at Harvard University, the
MEC/Fulbright Commission, and {\it ``Instituto de Astrof\'{\i}sica de 
Canarias''} and the {\it ``Deutsche Forschungsgemeinschaft''}.
This work has been partially supported by the Spanish DGES under projects 
PB95-1132-C02-01, PB98-0531-C02-02, and  {\it ``Plan Nacional del Espacio''}, 
under grant ESP98--1339-CO2. DBN and RGL appreciate the hospitality of the 
{\it ``Osservatorio Astronomico di Capodimonte''}.
We thank Alfredo Tripicchio for making available his data.
We do appreciate the very useful suggestions by the referee, Dr. Sofia
Randich.


\begin{appendix}{
Appendix A: the photometry}

In order to compute the observed photometry for a star covered with spots and
faculae, we have to estimate the individual contribution of each region. The
total luminosity can be expressed as:

\begin{equation}
{\rm 
  L_{\rm V}^{\rm obs} = L_{\rm V,spot} + L_{\rm V,faculae} + L_{\rm V,quiet} ,
}
\end{equation}

\noindent where L$_{\rm V}^{\rm obs}$, L$_{\rm V,spot}$, 
L$_{\rm V,faculae}$, and 
L$_{\rm V,quiet}$ 
are the luminosities of the observed star, and the contributions
from the spots, faculae and quiet region. We have assumed that the photometric
properties of the region covered with the faculae are the same as the quiet
photosphere. Since:

\begin{equation}
{\rm 
  L_{\rm V}/L_{\rm V,\odot} = 10^{-0.4(M_{\rm V}-M_\odot)} ,   
}
\end{equation}

\noindent we can conclude that:

\begin{equation}
{\rm  
  M_{\rm V}^{\rm obs} = 
-2.5\log\{10^{-0.4\times~M_{\rm  V,spot}}
+10^{-0.4\times~M_{\rm  V,faculae} }
+10^{-0.4\times~M_{\rm  V,quiet} }   \} , 
}
\end{equation}

\noindent where M$_{\rm V,spot}$, M$_{\rm V,faculae}$, 
and M$_{\rm  V,quiet}$ are the
magnitudes of the different regions. We can estimate the first value from:

\begin{equation}
{\rm 
  M_{\rm V,spot} = M_{\rm V,star}^{\rm (B-V)_{\rm spot}} -
 2.5\log~\{{\rm  L_{\rm V,spot}  \over  
L_{\rm V,star}^{\rm (B-V)_{\rm spot}} 
} \} ,
}
\end{equation}

\noindent where M$_{\rm V,star}^{\rm (B-V)_{\rm spot}}$  is the magnitude of a
star at the same (B--V) of the spot and  L$_{\rm V,star}^{\rm (B-V)_{\rm
spot}}$ its luminosity. The ratio between the two luminosities can be
re-written as:

\begin{equation}
 {\rm  L_{\rm V,spot}  \over  L_{\rm V,star}^{\rm (B-V)_{\rm spot}}  } =
{\rm  S_{\rm spot}\times~f_{\rm V,spot}  \over  
4\Pi~R^2_{\rm (B-V)_{\rm spot}}\times~f_{\rm V,star}^{\rm (B-V)_{\rm spot}} , 
}  
\end{equation}

\noindent where S$_{\rm spot}$ is the surface covered with spots and  f$_{\rm
V,spot}$ is the surface flux of the same region,  R$_{\rm (B-V)_{\rm spot}}$ is
the radius of a star with the same color as the spot and  f$_{\rm V,star}^{\rm
(B-V)_{\rm spot}}$  is its surface flux. By definition, we have that  f$_{\rm
V,spot}$=f$_{\rm V,star}^{\rm (B-V)_{\rm spot}}$.  Defining the filling factor
as:

\begin{equation}
{\rm 
 F_{\rm spot} = {\rm   S_{\rm spot} \over 4\Pi~R^2_{\rm quiet}  },  
}
\end{equation}

\noindent we have:

\begin{equation}
  {\rm  L_{\rm V,spot}  \over  L_{\rm V,star}^{\rm (B-V)_{\rm spot}}}  =
  [{\rm R_{\rm quiet} \over R_{\rm (B-V)_{\rm spot}}  }]^2\times~F_{\rm spot} .
\end{equation}

And we have the magnitude of the region covered with spots:

\begin{equation}
{\rm 
 M_{\rm V,spot} = M_{\rm V,star}^{\rm (B-V)_{\rm spot}} 
- 2.5 \log
\{\rm F_{\rm spot}\times~(R_{\rm (B-V)_{\rm quiet}}/R_{\rm (B-V)_{\rm spot}})^2 \} . 
}
\end{equation}

The case of the quiet region and the faculae is simpler, since the 
correction factor depending on the radii does not exist. Therefore:

\begin{equation}
{\rm 
 M_{\rm V,faculae} = M_{\rm V,star
}^{\rm (B-V)_{\rm faculae}} - 2.5
\log F_{\rm faculae}    
}
\end{equation}

\begin{equation}
{\rm 
M_{\rm V,quiet} = M_{\rm V,star}^{\rm (B-V)_{\rm quiet}} 
- 2.5 \log \{\rm 1-F_{\rm spot}-F_{\rm faculae} \} . 
}
\end{equation}


\end{appendix}

\begin{appendix}{Appendix B: the equivalent width}

Given a star with a quiet region and faculae and spots on its surface, 
the observed equivalent width contains contributions from each region. 
The equivalent width is defined as:

 \begin{equation}
{\rm 
 W^{\rm line}_{\rm obs} =  \int {\rm f(\lambda)^{\rm cont}
 - f(\lambda_l)^{\rm line} \over
 f(\lambda)^{\rm cont}} d\lambda    ,
}
 \end{equation}

\noindent where  f$(\lambda)^{\rm cont}$ 
and  f$(\lambda)^{\rm line}$ are the continuum and line fluxes. 
Since:

\begin{equation}
{\rm 
f(\lambda)^{\rm cont} = F_{\rm spot}\times~f_{\rm spot}^{\rm cont}  + 
F_{\rm faculae}\times~f_{\rm faculae}^{\rm cont} +
F_{\rm quiet}\times~f_{\rm quiet}^{\rm cont}
}
\end{equation}

\begin{equation}
{\rm 
f(\lambda)^{\rm line} = F_{\rm spot}\times~f_{\rm spot}^{\rm line}  + 
F_{\rm faculae}\times~f_{\rm faculae}^{\rm line} +
F_{\rm quiet}\times~f_{\rm quiet}^{\rm line}
}
\end{equation}

We have that the equivalent width can be re-written as :

\begin{equation}
{\rm 
 W^{\rm line}_{\rm obs} = 
 \int {\rm  F_{\rm spot}~(f_{\rm spot}^{\rm cont}-f_{\rm spot}^{\rm line}) + 
 F_{\rm faculae}~(f_{\rm faculae}^{\rm cont}-f_{\rm faculae}^{\rm line}) +
 F_{\rm quiet}~(f_{\rm quiet}^{\rm cont}-f_{\rm quiet}^{\rm line})  \over
F_{\rm spot}~f_{\rm spot}^{\rm cont}  + 
F_{\rm faculae}~f_{\rm  faculae}^{\rm cont} +
F_{\rm quiet}~f_{\rm quiet}^{\rm cont}   } d\lambda .
}
\end{equation}

This last equation has three different components, which
can be evaluated independently. Therefore, we have
W$^{\rm line}_{\rm obs}$  = [1] + [2] + [3].

\begin{equation}
{\rm 
[1] = 
\int {\rm    F_{\rm spot}~(f_{\rm spot}^{\rm cont}-F_{\rm spot}^{\rm line})/f_{\rm spot}^{\rm cont}  
  \over F_{\rm spot} + F_{\rm faculae}~f_{\rm faculae}^{\rm cont}/f_{\rm spot}^{\rm cont} + 
F_{\rm quiet}~f_{\rm quiet}^{\rm cont}/f_{\rm spot}^{\rm cont}    } d\lambda .
}
\end{equation}

We can define  $\alpha$$_{\rm line}$ = 
f$_{\rm spot}^{\rm cont}$/f$_{\rm quiet/faculae}^{\rm cont}$, 
which represents the ratio between continuum fluxes from the spot and the
quiet photosphere. Then:

\begin{equation}
{\rm 
[1] = 
{\rm  \alpha_{\rm line}~F_{\rm spot}  \over
\alpha_{\rm line}~F_{\rm spot} + F_{\rm faculae} + 
F_{\rm quiet} }\times~W_{\rm spot}^{\rm line}.
}
\end{equation}

On the other hand:

\begin{equation}
{\rm 
[2] = 
\int {\rm  F_{\rm faculae}~(f_{\rm faculae}^{\rm cont}-f_{\rm faculae}^{\rm line})~/f_{\rm faculae}^{\rm cont}  
 \over 
(F_{\rm spot}~f_{\rm spot}^{\rm cont} + 
F_{\rm faculae}~f_{\rm faculae}^{\rm cont} +
 F_{\rm quiet}~f_{\rm quiet}^{\rm cont})/f_{\rm faculae}^{\rm cont}   } d\lambda .
}
\end{equation}

So

\begin{equation}
{\rm 
[2] = {\rm   F_{\rm faculae} \over 
 \alpha_{\rm line}~F_{\rm spot} + F_{\rm faculae} +
 F_{\rm quiet}  }\times~W_{\rm faculae}^{\rm line} .
}
\end{equation}

Finally:
 
\begin{equation}
{\rm 
[3] =
\int {\rm  F_{\rm quiet}~(f_{\rm quiet}^{\rm cont}-f_{\rm quiet}^{\rm line})~/f_{\rm faculae}^{\rm cont}  
 \over  
 (F_{\rm spot}~f_{\rm spot}^{\rm cont} + F_{\rm faculae}~f_{\rm faculae}^{\rm cont}
 + F_{\rm quiet}~f_{\rm quiet}^{\rm cont})/f_{\rm faculae}^{\rm cont}     } d\lambda .
}
\end{equation}

Which can be written as:

\begin{equation}
{\rm 
[3] =
{\rm     F_{\rm quiet} \over
 \alpha_{\rm line}~F_{\rm spot} + F_{\rm faculae} + 
F_{\rm quiet} }\times~W_{\rm quiet}^{\rm line} .
}
\end{equation}

Now, we can evaluate the observed equivalent width:

\begin{equation}
{\rm 
 W_{\rm line}^{\rm obs} = {\rm 
\alpha_{\rm line}~F_{\rm spot}~W^{\rm line}_{\rm spot} +
F_{\rm faculae}~W^{\rm line}_{\rm faculae} +
F_{\rm quiet}~W^{\rm line}_{\rm quiet} \over \alpha_{\rm line}\times~F_{\rm 
spot}
+ F_{\rm faculae} + [1 - F_{\rm spot} - F_{\rm faculae}]     } .
}
\end{equation}

\end{appendix}


\end{document}